\documentclass[opre,nameyear]{informs4}
\usepackage{latexsym}
\usepackage{graphicx}
\usepackage[T1]{fontenc}

\usepackage{amsmath}
\usepackage{amsfonts}
\usepackage{amssymb}
\usepackage{amsbsy}
\usepackage{natbib}
\bibpunct[, ]{(}{)}{,}{a}{}{,}\def\bibfont{\footnotesize}

\usepackage[bookmarks=false,hypertexnames=false,colorlinks=true,urlcolor=blue,citecolor=blue,anchorcolor=blue,linkcolor=blue]{hyperref}

\usepackage{dsfont}

\usepackage{subfig}
\usepackage{xspace}
\newcommand{\RnS}{R\&S\xspace}

\newcommand{\EOC}{\textup{EOC}\xspace}
\newcommand{\PCS}{\textup{PCS}\xspace}
\newcommand{\PGS}{\textup{PGS}\xspace}
\newcommand{\PICS}{\textup{PICS}\xspace}

\newcommand{\EOCmax}{\Delta_{\max}}

\newcommand{\OCBA}{\textup{OCBA}\xspace}
\newcommand{\OCBAtwo}{\textup{OCBA}$_2$\xspace}

\newcommand{\indic}{\mathds{1}}

\newcommand{\cost}{c}           \newcommand{\costs}[1]{{\cost}_{#1}}

  \newcommand{\X}{X}      \newcommand{\Xs}[1]{{\X}_{#1}}
\newcommand{\Xvecs}[1]{\X_1, \cdots, \X_{#1}}

\newcommand{\costvec}{\boldsymbol c}

\newcommand{\nvec}{\boldsymbol n}
\newcommand{\thetavec}{\boldsymbol\theta}

\newcommand{\Svecsq}{\boldsymbol S^2}

\newcommand{\vecsigsq}{\boldsymbol\sigma^2}

\newcommand{\vecsigsqbnd}{\overline{\boldsymbol\sigma}^2}
\newcommand{\vecsigmabnd}{\overline{\boldsymbol\sigma}}

\newcommand{\predvars}[1]{\tilde{\sigma}^2_{#1}}
\newcommand{\predstds}[1]{\tilde{\sigma}_{#1}}

\newcommand{\mean}{\theta} 
\newcommand{\prmean}{\mu}   \newcommand{\prmeans}[1]{{\prmean}_{#1}}
       \newcommand{\vars}[1]{\sigma^2_{#1}}
\newcommand{\stds}[1]{\sigma_{#1}}
   
 \newcommand{\postvars}[1]{\sigma^2_{#1}}  \newcommand{\sampmeans}[1]{\bar{\X}_{#1}}

\renewcommand{\eqref}[1]{(\ref{#1})} 

\newcommand{\figref}[1]{Figure~\ref{#1}}

\newcommand{\tabref}[1]{Table~\ref{#1}}
\newcommand{\lemref}[1]{Lemma~\ref{#1}}

\newcommand{\propref}[1]{Proposition~\ref{#1}}

\newcommand{\thmref}[1]{Theorem~\ref{#1}}

\newcommand{\corref}[1]{Corollary~\ref{#1}}
\newcommand{\secref}[1]{Section~\ref{#1}}
\newcommand{\secrefb}[2]{Sections~\ref{#1} and \ref{#2}}

\newcommand{\appref}[1]{Appendix~\ref{#1}}

\newcommand{\Normal}{\mbox{Normal\xspace}}
\newcommand{\E}{\mathrm{E}}

\newcommand{\Var}{\mathrm{Var}}

\newcommand{\EOS}{EoS\xspace}

\newcommand{\stoptime}{\tau}
\newcommand{\stoptimeopt}{\stoptime^\ast}

\newcommand{\numproj}{k}

\newcommand{\myabs}[1]{\left| #1 \right|}

\newcommand{\filtration}[1]{\mathcal{F}_{#1}}
\newcommand{\OCs}[1]{\mbox{OC}_{#1}}

\newcommand{\contin}{\mathcal{C}}
\newcommand{\nreps}[1]{n_{#1}}

\newcommand{\EOSonekone}{EoS$_1$\xspace}
\newcommand{\EOSonektwo}{EoS$_1$\xspace} 
 \newcommand{\twoseosvarbound}{EoS$_2$Cap\xspace}
\newcommand{\twoseosvarplugin}{EoS$_2$Plugin\xspace}
\newcommand{\twoseosvarconfidence}{EoS$_2$Alpha\xspace}
\newcommand{\seqeos}{EoS-Seq\xspace}

\newcommand{\budget}{B}
\newcommand{\alphavar}{\alpha_\sigma}           \newcommand{\alphavars}[1]{\alpha_{\sigma,#1}}

\newtheorem{theorem}{Theorem}
\newtheorem{lemma}[theorem]{Lemma}
\newtheorem{proposition}[theorem]{Proposition}
\newtheorem{corollary}[theorem]{Corollary}

\newtheorem{remark}{Remark}
\renewcommand{\halmos}{\hfill\ensuremath{\square}}

\begin{document}

\RUNAUTHOR{Chick, Li, and Hong}
\RUNTITLE{Economics-of-Selection R\&S Balances Sampling Cost and Selection Quality}
\newcommand{\papertitle}
{Economics of Selection in Ranking \& Selection: Neither Fixed-Precision nor Fixed-Budget}
\TITLE{\papertitle}
\ARTICLEAUTHORS{
    \AUTHOR{Stephen E. Chick}
    \AFF{Technology and Operations Management Area, INSEAD, Fontainebleau, France\\\EMAIL{stephen.chick@insead.edu}}

    \AUTHOR{Zaile Li}
    \AFF{Department of Industrial Systems Engineering and Management, National University of Singapore, Singapore \\\EMAIL{zaile.li@nus.edu.sg}}

    \AUTHOR{L. Jeff Hong}
    \AFF{Department of Industrial and Systems Engineering, University of Minnesota, Minneapolis, MN, USA\\\EMAIL{lhong@umn.edu}}
}
\ABSTRACT{Ranking and selection (R\&S) has evolved tremendously over the last several decades, with applications in agriculture, discrete optimization with computer simulation, clinical trial design, and other areas.
We challenge a widely accepted classification of R\&S procedures into two categories, fixed precision and fixed budget.
We discuss a third class, economics of selection (\EOS) procedures, which is conceptually distinct and is found in extant literature on Bayesian hypothesis testing and simulation optimization.
Procedures in this third class seek to optimally balance two quantities: the cost of learning about the performance of each alternative and the quality of the alternative that is ultimately selected on the basis of that learning.  Fixed precision and fixed budget procedures tend to optimize one of these quantities, while putting a constraint on the other. We clarify these differences and show why it matters.
We then present new frequentist R\&S procedures in this third category, including both two-stage procedures and a sequential heuristic, assuming Gaussian observations that have either known or unknown variances.
We find that \EOS is not a question of Bayesian or frequentist, nor of asymptotic or finite sample analysis, nor of finite-stage or fully sequential,
nor of one mean versus a standard or comparing multiple alternatives.
We identify opportunities for further R\&S research.
}\KEYWORDS{ranking and selection; best arm identification; simulation optimization; expected opportunity cost; Bayesian procedures; frequentist procedures}
\HISTORY{\today}
\maketitle

\section{Introduction}
\label{sec:intro}
Ranking and selection (\RnS) is a branch of simulation optimization that studies the selection of the ``best'' of a finite set of simulated alternatives, where ``best'' is defined with respect to selecting the alternative with the largest mean. The mean of each alternative is unknown but can be inferred with statistical sampling \citep{goldsman:selection94,goldsman:selection98,bechhofer:selection,Kim2006,chick:chapter04}. \RnS is also used in other important fields, including agriculture and medicine, and with variations on the formulation: best arm identification (BAI), pick a subset of alternatives better than a threshold, identify the top $m$ alternatives, and so forth.

Central to \RnS procedures are allocation rules, stopping rules, and selection rules. Allocation rules determine which alternative or alternatives get sampled next. Stopping rules determine when the sampling should stop so that the best can be selected. Selection rules determine, on the basis of observed data, which alternative to select as best. Allocation rules are sometimes called sampling rules, and selection rules are sometimes called implementation decisions or decision rules \citep{ryzhov:information,garivierkaufmann16fixedconf}.
There are both frequentist and Bayesian approaches to allocation, stopping, and selection. Allocations have included deterministic rules, rules with a small number of stages, and highly sequential response-adaptive allocations.
Stopping rules have included deterministic or response-adaptive approaches, too.
The quality of selection decisions can be measured in different ways, such as with the probability of correct selection (\PCS), probability of good selection (\PGS), or the expected opportunity cost (\EOC) of a potentially incorrect selection (also called simple regret). \PCS and \PGS are the probabilities that the alternative selected by a procedure is the true best, or close to best, respectively. And \EOC is the average difference in reward from the selected alternative and that from the true best.

With so many variations, there is understandably great interest in classifying the huge range of potential options. Classifications can help modelers select an \RnS procedure that is suitable for a given task, understand the impact of hidden assumptions that affect inference, and find linkages that open new research questions.

Several reviews provide useful guidance for such classifications.
\citet{hunternelson} provide three classifications that distinguish \textit{fixed-precision} versus \textit{fixed-budget} guarantees, finite-sample versus asymptotic validity, and frequentist versus Bayesian guarantees. They acknowledge that the classification boundaries of these dichotomies are not always sharp.
\citet{hongfanluo} focus on the importance of the fixed-precision versus fixed-budget perspective rather than the frequentist versus Bayesian perspective, noting the links of fixed-precision to hypothesis testing and of fixed-budget to dynamic programming.
\citet{HONG2026701} repeat that focus and state ``Generally, there are two formulations of the \RnS problem: the fixed-precision formulation and the fixed-budget formulation.''  The BAI literature also tends to classify procedures as fixed-precision or fixed-budget
\citep{garivierkaufmann16fixedconf}.

These important reviews all overlook a third class of \RnS procedures: \textit{economics of selection} procedures.
This class is not new in general (e.g., \citealp{chernoff:sequential61,bather:optimal70,degroot:decisions}) nor in simulation (e.g., \citealp{chick:09a,chick:12a}) or clinical trials and health economics (e.g., \citealp{chick2017bayesian,ENACTposition}).

\citet{chick:wscselect05} may have unintentionally contributed to the oversight. They classified three types of \RnS procedures in simulation: optimal computing budget allocation (\OCBA, \citealp{chenhc:lowerbound,chenhc:selection97}), expected value of information (EVI, \citealp{Chick2001}), and indifference zone (IZ, \citealp{Kim2006}).
EVI gained visibility as a Bayesian approach for allocation rules. This might be because many explored \OCBA and EVI for allocation rules with fixed budgets. Groundbreaking work on the knowledge gradient (KG, \citealp{frazier2008,frazier2010paradoxes,ryzhov:information}) derives important results for Bayesian EVI-type allocation rules with either a finite or asymptotically large fixed budget.
There was less attention to EVI's decision-theoretic potential for new stopping rules. It is understandable why some might classify EVI as fixed-budget and not see its role for a third class of \RnS via stopping rules.

Economics of selection (\EOS) procedures optimize the expected profit of a selection project: they balance the monetary cost of sampling resources up to the stopping time plus the financial reward of a selection decision.
Fixed-budget focuses on resource consumption for sampling, not profit.
Fixed-precision focuses on achieving a selection quality goal, usually from a hypothesis testing framework, not profit.
Fundamentally, \EOS differs in that it considers the cost of an inferior selection decision (an opportunity cost from not having perfect information) and the cost of sampling (due to allocation rule choices up to a stopping time) to be measurable on a common scale, and \textit{the total cost of sampling and opportunity cost of inferior decisions}, in expectation, is minimized.
Stopping times involve a balance: larger sample sizes cost more but give more expected value of sample information for correct selections.

We show how \EOS procedures differ from fixed-precision and fixed-budget procedures, discuss the value of considering them as a third category, and argue that \EOS procedures may be more relevant than those other two categories for decision makers who care about the total financial value of the selection decision, net of the cost of analysis.
To that end, we clarify and further distinguish these three approaches in \secref{sec:threeclasses} and then give a simple worked example to illustrate important differences between them in \secref{sec:examples}.
We demonstrate the usefulness of the approach for creating new types of policies by
new frequentist \EOS procedures in \secref{sec:freqeos} for one- and two-stage procedures with multiple alternatives and potentially unknown variances, and in \secref{sec:freqeosseq} for sequential \EOS heuristics with a finite set of alternatives. These new frequentist \EOS procedures advance earlier work \citep{threeflavorswsc26} that considered one-stage procedures with one or two arms with unknown means and known variances.
We explore these procedures numerically in \secref{sec:numericalstudy}. These sections also identify areas of future research for Bayesian and frequentist \EOS \RnS procedures, as do our conclusions in \secref{sec:discuss}.

\section{Three Classes of \RnS Procedures}
\label{sec:threeclasses}

We describe and distinguish fixed-precision procedures (no budget constraint, with selection-error guarantees), fixed-budget procedures (with budget constraint, without selection-error guarantees), and economics of selection procedures (neither budget constraint nor selection-error guarantee, but a balance between sampling costs and loss in value from potentially incorrect selections).

\subsection{Fixed-precision \RnS Procedures}

Fixed-precision procedures seek to satisfy a constraint on the quality of the selection, possibly subject to constraints. Many frequentist indifference-zone (IZ) procedures are designed to guarantee that the $\PCS \geq 1-\alpha$ whenever the true mean of the best is at least $\delta$ better than the means of all other alternatives, with as small a sampling effort as possible \citep{bechhofer1954single,bechhofer:selection,kim2001fully}. \citet{hunternelson} allow for a broad interpretation of constraints on the quality of a solution, such as constraints on the probability of a good selection (within $\delta$ of best), or the probability for correct subset selection, and other generalizations.

We fully support the notion of flexibility in defining statistical constraints on selection quality.
For example, \citet{eckman2021fixed} advocate for the probability of good selection (\PGS), or of selecting an alternative within $\delta$ of the best.
The constraint for fixed precision procedures can also be with respect to \EOC. \citet{chick:izopcost} give two two-stage frequentist procedures that guarantee the \EOC is no greater than a prespecified upper bound, $\EOCmax$.

In fixed precision procedures, ``good'' is often taken to mean a smaller amount of resources used for sampling to achieve the precision goal.  \citet{hunternelson} also allow for a broad interpretation of resource usage, such as sample size, computing time, or clock time. It might also be the monetary cost of getting data.
For deterministic stopping rules, a fixed sample size is used. For response adaptive stopping rules, the expected amount of resources is to be minimized. Constraints can be expressed as a bound on the expectation of a loss function. For example, the frequentist $\PICS$ is the expectation of a loss function that is 0 if the best alternative is correctly selected, and 1 if an incorrect selection is made.

Qualitatively, ``good'' fixed-precision procedures solve the following optimization:
\begin{equation}
\label{def:fixedprecision}
\begin{array}{rl}
    \mbox{minimize}  & \mbox{sampling resources (e.g., clock time, sample size, computing budget)} \\
    \mbox{such that} & \mbox{expected loss from selection is bounded (e.g., \PICS $\leq \alpha$ or \EOC $\leq \EOCmax$)} \\
     & \mbox{for all problem configurations in a set} \\
\end{array}
\end{equation}

The statement ``for all problem configurations in a set'' merits discussion. For frequentist procedures, the expected loss target is often a worst-case bound constraint that holds for every problem configuration in the set.  With the IZ approach and normally distributed outcomes, this is usually the set of problem configurations where the mean of the ``best'' alternative is at least some prespecified amount $\delta > 0$ better than the mean of all of the others.
Some recent breakthroughs allow for indifference-zone-free frequentist guarantees \citep{FanHongNelson2016izfree}, with $\delta = 0$.
For Bayesian procedures, the loss constraint holds on average, the expectation taken over the prior distribution over the unknown problem configurations of interest, and $\delta$ is essentially assumed to be 0.

In the Bayesian \RnS literature,
\citet{brankechickschmidt:selection07} study a wide range of old and new sequential allocation and stopping rules, including some with Bayesian posterior \PICS and \EOC guarantees.
\citet{EckHenderson2022} give novel techniques that improve the computation of such rules and others that provide Bayesian fixed-precision stopping times. Thus, not all Bayesian procedures are fixed-budget. Such procedures have Bayesian guarantees but often not frequentist ones. \citet{frazierBIZ} gives one notable exception of a Bayesian procedure with frequentist guarantees.
In the BAI literature, \cite{garivierkaufmann16fixedconf} present stopping rules for \RnS procedures that provide frequentist \PCS guarantees for a broad class of allocation rules, assuming $\delta > 0$ and observations are from the one-parameter exponential family of distributions.

\subsection{Fixed-budget \RnS Procedures}

Fixed-budget \RnS procedures focus on minimizing the expected loss due to a potentially incorrect selection of the best alternative, subject to a sampling resource constraint. That constrained resource may again be measured in a variety of ways, such as total sample size, total computing time, total financial costs of obtaining samples, and so forth. The quality may also be measured with respect to different metrics, such as minimizing a proxy for \PICS \citep[typical for \OCBA]{chenhc:gradalloc96}, or the EVI and KG procedures' common focus on minimizing \EOC.
In optimization terms, fixed-budget can be thought of as a satisfying version of fixed-precision.

Qualitatively, ``good'' fixed-budget procedures solve the following optimization:
\begin{equation}
\label{def:fixedbudget}
\begin{array}{rl}
    \mbox{minimize} & \E [\mbox{loss from selection}] \mbox{ (e.g., \PICS  or \EOC)} \\
    \mbox{such that} & \mbox{sampling resources are bounded  (e.g., clock time, sample size, computing budget)} \\
     & \mbox{for all problem configurations in a set} \\
\end{array}
\end{equation}

Much of the \OCBA literature and KG literature studies settings where there is a fixed computing budget. The EVI literature has often studied fixed budget examples, too, though, as we noted above, there have been fixed-precision algorithms for EVI as well. These approaches are Bayesian or decision theoretic, and the fixed-budget class is sometimes associated with Bayesian procedures.

There also exist frequentist fixed-budget procedures. We provide some examples. Under the IZ formulation,  \citet{bechhofer1954single} derive a one-to-one correspondence between the precision parameter $\alpha$ and the total sample size $N$ for the classical case with known and equal variances. Consequently, a fixed-precision procedure can be readily converted into a fixed-budget one by determining the corresponding $\alpha$ for a given budget $N$. Large-deviation results have given insights into asymptotic behavior of selection procedures \citep{glynn2004large,ryzhov22LD}.
Some interesting links between the \EOC and \PICS have been established \citep{ryzhov16rates,gao2017new}, including for asymptotically large budget allocations.
For settings involving many alternatives, \citet{hong2022solving} and \citet{li2025surprising} develop frequentist fixed-budget procedures that achieve optimal rate guarantees on the sampling budget required for a nonvanishing precision level.

\subsection{Economics of Selection \RnS Procedures}
\label{sec:econseldef}

Importantly, classifying \RnS into only fixed-precision or fixed-budget procedures does not fully account for a central idea that drives organizational decision-making: money, or more generally, project utility.
Sampling costs in clinical trials can be tens of thousands of dollars per patient observation. Computing costs for simulation observations typically cost less, but still use CPU time, energy, and clock time. The costs of computation are becoming more salient with the ``tokenomics'' of generative AI products.
Sampling budgets are typically expressed in terms that can be converted to money, hence the term ``budget''. If the simulation project takes months, then discounting may be relevant. We focus now on undiscounted costs and rewards.

Moreover, decisions made based on \RnS procedures have financial value in many applications. Even if a simulation \RnS problem is formally stated ``Which alternative maximizes long-run production capacity?', it is the monetary value of the decision that is important to a firm. In clinical trials, the question ``which treatment is better?'' has implications for drug financial costs, for the reduction of costly complications, as well as social and health issues. For example, \citet{2025nicethresholdupdate} uses conversions of quality-adjusted life year (QALY) gains that a new drug might bring relative to the existing standard of care to inform treatment adoption decisions for England.

Thus, there is a third class of \RnS procedures, beyond fixed-precision and fixed-budget: economics of selection (\EOS). More sampling means greater sampling costs. More sampling also provides the potential to make a more informed decision and lower expected losses from potentially incorrect selections.
\EOS procedures minimize the expected penalty accounting for both of these costs, unlike procedures which optimize just one or the other. \textit{There is no fixed budget nor fixed precision or selection quality target} that is specified up front.
We avoid questions like ``What $\alpha$ should I use for a fixed-confidence procedure?'', ``Should the choice of $\alpha$ be the same whether observations cost \$0.0001 or \$100,000?'', and ``What sampling budget is right for the selection decision at hand?''
The analysis below shows how an appropriate value of $\alpha$ or sampling budget naturally and endogenously emerges from an \EOS procedure.

\citet{chernoff:sequential61,bather:optimal70} and \citet{degroot:decisions} used the term ``Bayesian decision-theoretic'' hypothesis testing for such procedures.  \citet{chick:09a} and \citet{chick:12a} coined the term ``economics of selection'' in the simulation community. ``Value-based clinical trial design'' has been used in clinical trials and health economics \citep{chick2017bayesian,ENACTposition}. Each of these works uses Bayesian models. We will show that \EOS designs may also be frequentist.

Qualitatively, ``good'' \EOS procedures solve the following optimization for expected total cost:
\begin{equation}
\label{def:econselect}
\begin{array}{rl}
    \mbox{minimize} & \E \mbox{[sampling resources + loss from selection] } \mbox{ (e.g. money, \EOC, QALYs)} \\
     & \mbox{for all problem configurations in a set} \\
\end{array}
\end{equation}

In \eqref{def:econselect}, adding sampling and selection costs (e.g., opportunity costs) requires them to have the same unit of measure. For simulations used for \RnS in commercial applications, it is natural for that common unit of measure to be money. For the clinical trial setting mentioned above, one might use money as the common unit of measure through cost/QALY conversions mentioned above. Alternatively, one might use health outcomes as a common measure, with the number of patients receiving a suboptimal treatment during the clinical trial being balanced with the expected number of patients that might receive a suboptimal treatment selected for broad usage on the basis of trial data. This statement is not to advocate the use of one measure over the other -- only to posit there are multiple ways to balance the expected costs of sampling and potentially incorrect decisions.

One might consider more general utility functions of sampling and selection than the linear combination in \eqref{def:econselect}. For example, if selection rewards are discounted and larger sample sizes delay selection times, \eqref{def:econselect} minimizes the sum of expected discounted sampling costs plus a selection loss measured with respect to an oracle who knows all parameters at time 0.

For Bayesian \EOS procedures, expectations are over prior distributions for the unknown configuration.  For frequentist \EOS procedures, we minimize the expected total cost of sampling plus the opportunity cost of selection, conditional on the configuration, as in \secrefb{sec:freqeos}{sec:freqeosseq}.

\subsection{Summary of These Three Approaches}

In summary, fixed budget is about improving the value of the selection decision subject to a resource consumption constraint. Fixed precision is about minimizing budget consumption to achieve a given statistical selection quality guarantee.  Economics of selection is about maximizing the financial value of an entire project, or equivalently, minimizing the expected sampling cost plus opportunity cost of selection with an \RnS procedure relative to an omniscient oracle at time $0$.

These approaches are interrelated. One might, for example, try to optimize the budget of a fixed-budget procedure to achieve a given fixed-precision goal or minimum total cost for \EOS. However, these three approaches remain distinct in their modeling goals, structure, and analysis.

\section{Examples of Bayesian Economics of Selection (\EOS) Procedures}
\label{sec:examples}

We now recall examples of Bayesian \EOS procedures with fixed sample sizes and with response-adapted sample sizes.
We demonstrate that they are neither fixed-precision nor fixed-budget and why it matters. To focus on key concepts, we do this for a simplified \RnS problem that will also be studied from a frequentist perspective in \secref{sec:freqeos}.

We hope to identify which of the $\numproj = 2$ alternatives has the larger mean. Alternative~1 has an unknown mean $\mean$ that can be inferred from simulation observations $\Xs{i}$ that have a known variance, $\sigma^2 = \Var[\Xs{i}]$. Alternative $2$ has a known deterministic mean $0$.
If $\mean$ is at least 0, then alternative 1 is best. Otherwise, alternative 2 is best. The stopping rule specifies the sample size, $\stoptime$, before selection occurs.
We consider both one-shot procedures, with a fixed $\stoptime$ chosen at time zero, and fully sequential procedures, where $\stoptime$ is a response-adaptive stopping rule, as described below.

The selection rule of the Bayesian procedures here uses the posterior mean at the time of stopping.
We assume that $\mean$ has a  $\Normal(\prmeans{0},\postvars{0})$ prior distribution.
We denote the effective number of observations in the prior distribution by $\nreps{0} = \sigma^2 / \postvars{0}$.
We let $\Psi(z) = \phi(z) - z \Phi(-z)$ denote the newsvendor loss function with \Normal(0,1) demand and we recall $d\Psi(z)/dz = \Phi(z)-1$ and $\Psi(-z)
= z + \Psi(z)$.
Thus, the expected loss with no sampling, when $\prmeans{0}\geq 0$, is
\begin{equation}
    \label{eq:lossattime0}
    \E [- \theta \indic(\theta < 0)] = \sqrt{\stds{0}} \Psi(\myabs{\prmeans{0}} / \sqrt{\stds{0}}).
\end{equation}
Standard results show that $\Pr(\mean \mid \Xvecs{\stoptime})
\sim \Normal(\prmeans{\stoptime}, \postvars{\stoptime})$, where $\prmeans{\stoptime} = (\nreps{0} \prmeans{0} + \stoptime \sampmeans{\stoptime})/(\nreps{0} + \stoptime)$ is the posterior mean and $\postvars{\stoptime} = \sigma^2 / (\nreps{0} + \stoptime)$ is the posterior variance of $\mean$ \citep{degroot:decisions}. We write $\nreps{\stoptime}= \nreps{0} + \stoptime$ for the effective sample size of this posterior distribution.

For our running example, we assume a sampling cost $\cost=1$ and output standard deviation $\sigma = 10^5$, unless otherwise noted.  We assume $\mean \sim \Normal (0, \sigma^2 / 5)$, so $\prmeans{0}=0$ and $\nreps{0} = 5$.

\subsection{One-Stage \EOS Procedure} \label{sub:Bayesdetexample}

\citet{degroot:decisions} discusses a Bayesian hypothesis test of whether the mean of a normal distribution is positive or not. It is the same as our economics of selection problem adapted to our $\numproj = 2$ setup, assuming the Bayesian model above.
We draw upon \citet[Sec.~11.8--11.9]{degroot:decisions} here.

Consider a fixed sample size $\stoptime$. The posterior mean $\prmeans{\stoptime}$ is not known at time 0 because $\Xvecs{\stoptime}$ has not been observed. We therefore look at the probability that $\prmeans{\stoptime}$, given prior information, will be positive or not. The predictive distribution of $\prmeans{\stoptime}$ is
$$
\Pr(\prmeans{\stoptime} \mid \prmeans{0})
    \sim \Normal\left( \prmeans{0} , \stoptime \sigma^2 / [\nreps{0} (\nreps{0} + \stoptime)] \right).
$$
We denote the predictive variance of $\prmeans{\stoptime}$ by $\predvars{\stoptime} = \stoptime \sigma^2 / [\nreps{0} (\nreps{0} + \stoptime)]$.

Suppose for the moment that $\prmeans{0} \geq 0$. Without observations, alternative 1 is selected as best. Observing $\stoptime$ outcomes changes that decision when $\prmeans{\stoptime} < 0$. The expected selection gain from changing decision is $\E[ -\prmeans{\stoptime}  \indic(\prmeans{\stoptime} < 0) ]$.
As $\stoptime \rightarrow \infty$, $\predvars{\stoptime} \rightarrow \vars{0}$ and the posterior mean converges to the true mean, $\prmeans{\stoptime}\rightarrow\mean$ with probability 1 by the Bayesian LLN. So, with perfect information, the expected selection gain is as in \eqref{eq:lossattime0}.
Each observation costs $\cost$, so the sampling resource cost is $\cost \stoptime$.
The minimum expected total sampling cost plus opportunity cost (OC) of incorrect selection in \secref{sec:econseldef} is thus \citep[eq.~(5) of Sec.~11.9]{degroot:decisions}:
\begin{equation}
\label{def:econselectBayeDeterministic}
\begin{array}{rl}
  \min_{\stoptime\geq 0} & \cost \stoptime
    + \sqrt{\stds{0}} \Psi (  \myabs{\prmeans{0} } / \sqrt{\stds{0}}))
    - \sqrt{\predvars{\stoptime}} \Psi (  \myabs{\prmeans{0} } / \sqrt{\predvars{\stoptime}}) \\
\end{array}
\end{equation}
It can be shown that the third term strictly and continuously converges to zero as $\stoptime$ drops to 0 from above (i.e., there is no gain if there is no new information) and that the expected OC, the sum of the second and third terms, decreases to 0 as $\stoptime \rightarrow \infty$ (i.e., positive expected value of information, and there is zero expected loss with perfect information).
There is therefore an optimal $\stoptime = \stoptimeopt$.

\begin{figure}[ht]
    \centering
    \subfloat[Total Cost is Minimized Near $\stoptime^* = 211$ (with $\cost=1$).]{
    \includegraphics[width=0.46\linewidth]{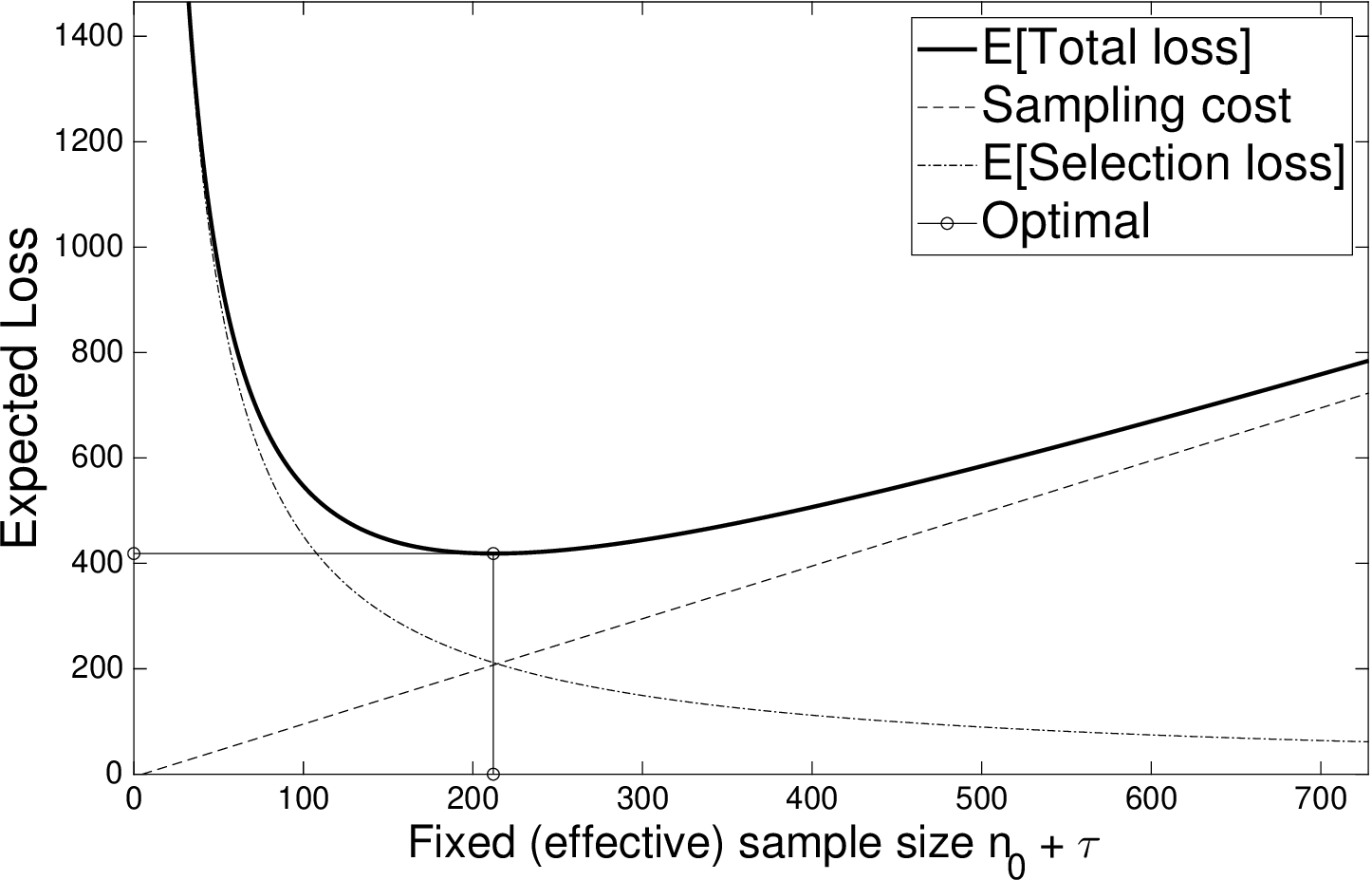}}
    \hfill
    \subfloat[Total Cost is Minimized Near $\stoptime^* = 595$ (with $\cost=1/8$).]{
    \includegraphics[width=0.46\linewidth]{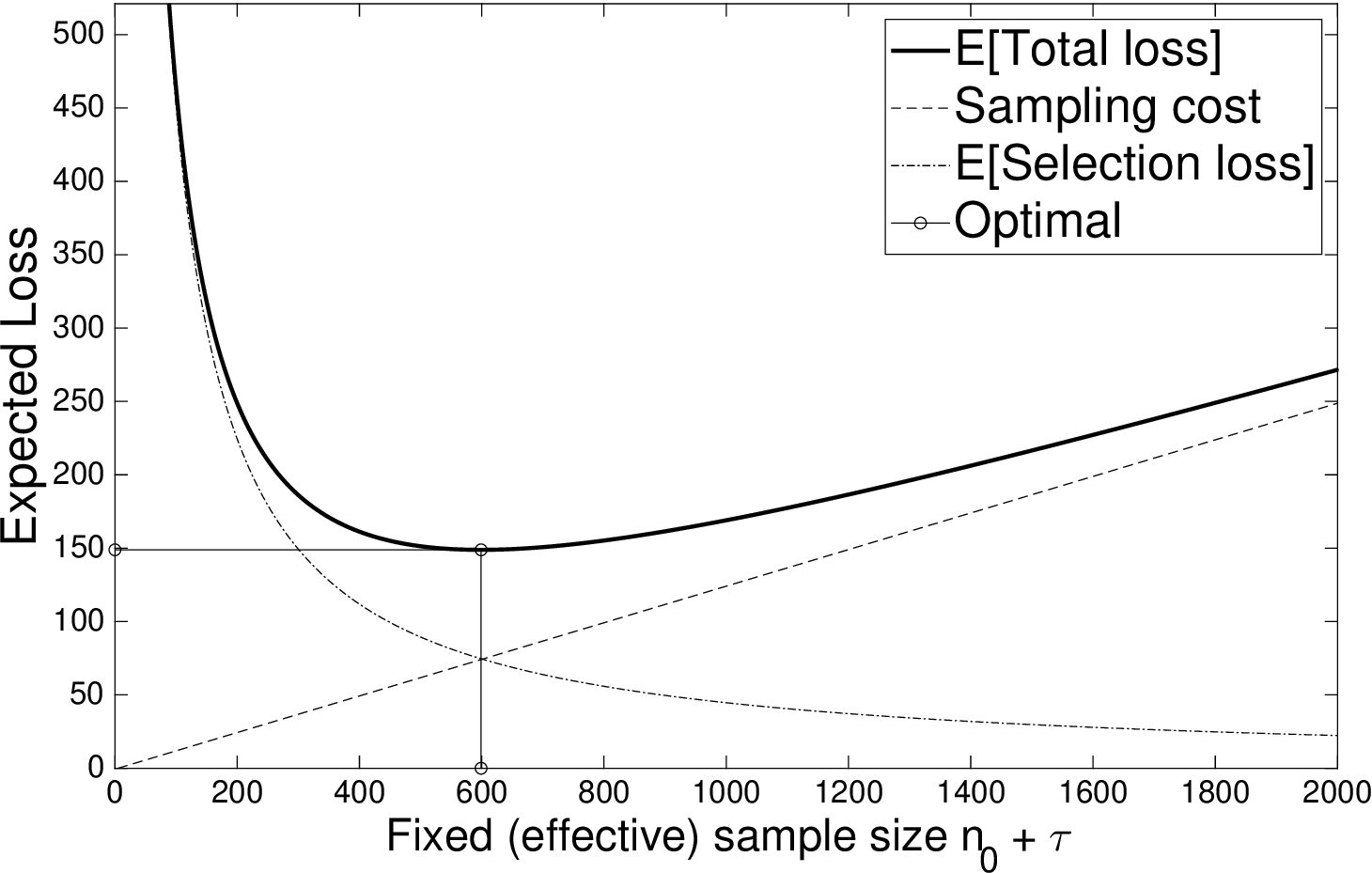}}
    \caption{Bayesian Fixed Sample Size: Sampling Costs Increase and E[Selection Costs] Decrease in Sample Size $\stoptime$ ($\sigma=10^5$, $\prmeans{0}=0$, $\nreps{0}=5$).}
    \label{fig:degrootexample}
\end{figure}

This is \textit{not a fixed-budget procedure}.
Sample size emerges from the variance and sampling cost, not from an up-front budget constraint.
\figref{fig:degrootexample} illustrates the sampling costs, expected selection costs, and expected total cost of \eqref{def:econselectBayeDeterministic} as a function of $\stoptime$.
For this example, with $\cost=1, \sigma=10^5$, $\mu_0 = 0$, $\nreps{0}=5$, we find $\stoptime^* = 207$ and a worst-case expected total cost in \eqref{def:econselectBayeDeterministic} of 418.6 with that $\stoptime$.
With $\cost = 1/8$, we find $\stoptimeopt = 594$ and a worst-case expected loss in \eqref{def:econselectBayeDeterministic} of 148.9.

This is \textit{not a fixed-precision procedure}.
There is no IZ parameter $\delta > 0$ nor an incorrect selection probability constraint, $\PICS \leq \alpha$. We report some statistics for comparison for the case with $c=1, \stoptimeopt=207$. The posterior \PICS, conditional on the true mean being $\delta=2,000$, for example, is $P(\prmeans{\stoptimeopt} < 0 \mid \mean = \delta) = 0.3854$ and the probability the sample average gives an incorrect selection is $P(\sampmeans{\stoptimeopt} < 0 \mid \mean = \delta) = 0.3868$. Both are worse than a \PICS of $\alpha = 0.05$. We note that such choices for $\delta,\alpha$ are arbitrary: the Bayes average case \PICS would be better than a worst case of 0.05 if $\delta$ had been chosen to be larger than $\Phi^{-1}(1-\alpha) \sigma / \sqrt{\nreps{\stoptimeopt}} \approx 11,405$, assuming $c=1$. And an $
\alpha=0.05$ \PICS guarantee with $\delta = 2000$ could have been obtained if the sample size were sufficiently large, which happens when sampling costs are sufficiently small, $\cost \in (0, 1/1,030)$.
This gives a sense of how financial parameters (cost of sampling, expected selection loss) link to statistical parameters.

Although \eqref{def:econselectBayeDeterministic} has terms that resemble the KG-beta factor that is useful in Bayesian sequential \RnS \citep{chickfrazier:wsc09,frazier2010paradoxes}, they are not the same. The KG$_\beta$ factor for the $\beta$-step look-ahead at time $t=0$ is $\sqrt{\predstds{\beta}} \Psi (  \myabs{\prmeans{0} } / \sqrt{\predstds{\beta}}) / (\cost \beta)$ and measures the rate of information accumulation per unit of sampling resource.

\subsection{Fully Sequential \EOS Procedure with Response-Adaptive Stopping Times}
\label{sub:Bayesseqexample}

\citet{chernoff:sequential61} introduces a Bayesian, decision-theoretic framework to test whether the unknown mean of a normal distribution is positive or not. The objective is to minimize the expected cost of sampling plus expected regret, where regret is the linear loss when an incorrect decision is made. That framing is essentially the economics of selection framing.
He allowed an analyst to stop sampling and decide the mean's sign after each observation or to continue, and triggered a stream of work (e.g.,  \citealt{chernoff:64a,chernoff:65a,chernoff:65b,bather:optimal70,bather:optimal83,lai:review2001}).

Their fully sequential techniques were developed to allow for $\numproj\geq 2$ alternatives for the simulation and clinical trials communities under the name economics of selection \RnS procedures.
\citet{chick:09a} optimized expected discounted rewards of the simulated alternative selected, less the discounted costs of sampling through the stopping time.
\citet{chick:12a} optimized undiscounted rewards from the simulated alternative selected, net sampling costs, as in this paper.
While sampling continues at times $t=0,1, \ldots$, the expected reward of stopping immediately is compared to the expected reward of obtaining one more observation and then having the option to continue optimally. If stopping is better, the stopping time is set, $\stoptime = t$, and an alternative is selected. If stopping is not better, then one more observation is taken, and the process repeats.
These were the first sequential \EOS procedures presented at WSC \citep{chickgans:wsc08,chickfrazier:wsc09,chick2015optimal}.
Computational solutions \citep{chernoff:petkau86} involve solving a free-boundary partial differential equation (PDE) for a heat equation, similar to pricing an American option on regular (not geometric) Brownian motion with unknown mean drift (or to computing dynamic-lookahead, instead of fixed-lookahead, KG factors).

In our example with $\numproj=2$ alternatives,
their model with undiscounted rewards is written:
\begin{equation}
    \label{eq:undiscountedstopprob}
    \max_{\tau\geq0} \E[- \cost \stoptime + \max\{0,\prmeans{\stoptime}\} |\filtration{0} ].
\end{equation}
The maximization is over non-anticipative stopping times adapted to the filtration $\filtration{t}$ determined by $\{\Xvecs{t}\}$.
We convert \eqref{eq:undiscountedstopprob} to the desired form of \eqref{def:econselect} by noting a) the true mean $\mean$ equals $\prmeans{\infty}$ almost surely when all alternatives are sampled infinitely often, b)  the expected loss from selection with linear loss at time $\stoptime$ is the expected value of perfect information (EVPI) minus the opportunity cost,
also called the linear loss function, or difference between true mean and the selected mean, as in \eqref{def:econselectBayeDeterministic} for fixed $\stoptime$,
and c) maximizing a function is the same as minimizing its negative plus a constant (the EVPI at time 0).
This gives an equivalent optimization in the desired form of \eqref{def:econselect}:
\begin{equation}
    \min_{\tau\geq0} \E[ \cost \stoptime + \OCs{\stoptime} |\filtration{0} ].
\end{equation}

They find the set $\contin$ of $(\nreps{\stoptime},\prmeans{\stoptime})$ where it is optimal to continue sampling and to stop otherwise.
 Let
$    \beta = c^{-1/3} \sigma^{-2/3}
    \mbox{ and }
    \gamma = c^{2/3} \sigma^{-2/3}.
$The quantity $\beta=(c \sigma^{2})^{-1/3}$ is the cube
root of the sampling efficiency
proportional to the product of the sampling variance and the cost to obtain the estimate.
One continues sampling as long as $\mu_t$, is in the interval
$  [-\beta^{-1} b(1/(\gamma \nreps{t})), \beta^{-1} b(1/(\gamma \nreps{t}))],
$ where $b()$ has an easily computed approximation that does not require a full solution of the free-boundary PDE that determines it \citep{chick:12a}.

 \begin{figure}[ht]
    \centering
    \subfloat[For posterior means between the dashed lines, it is optimal to continue; above and below it is optimal to stop. Different sample paths leave the continuation set at different times.]{\includegraphics[width=0.48\linewidth]{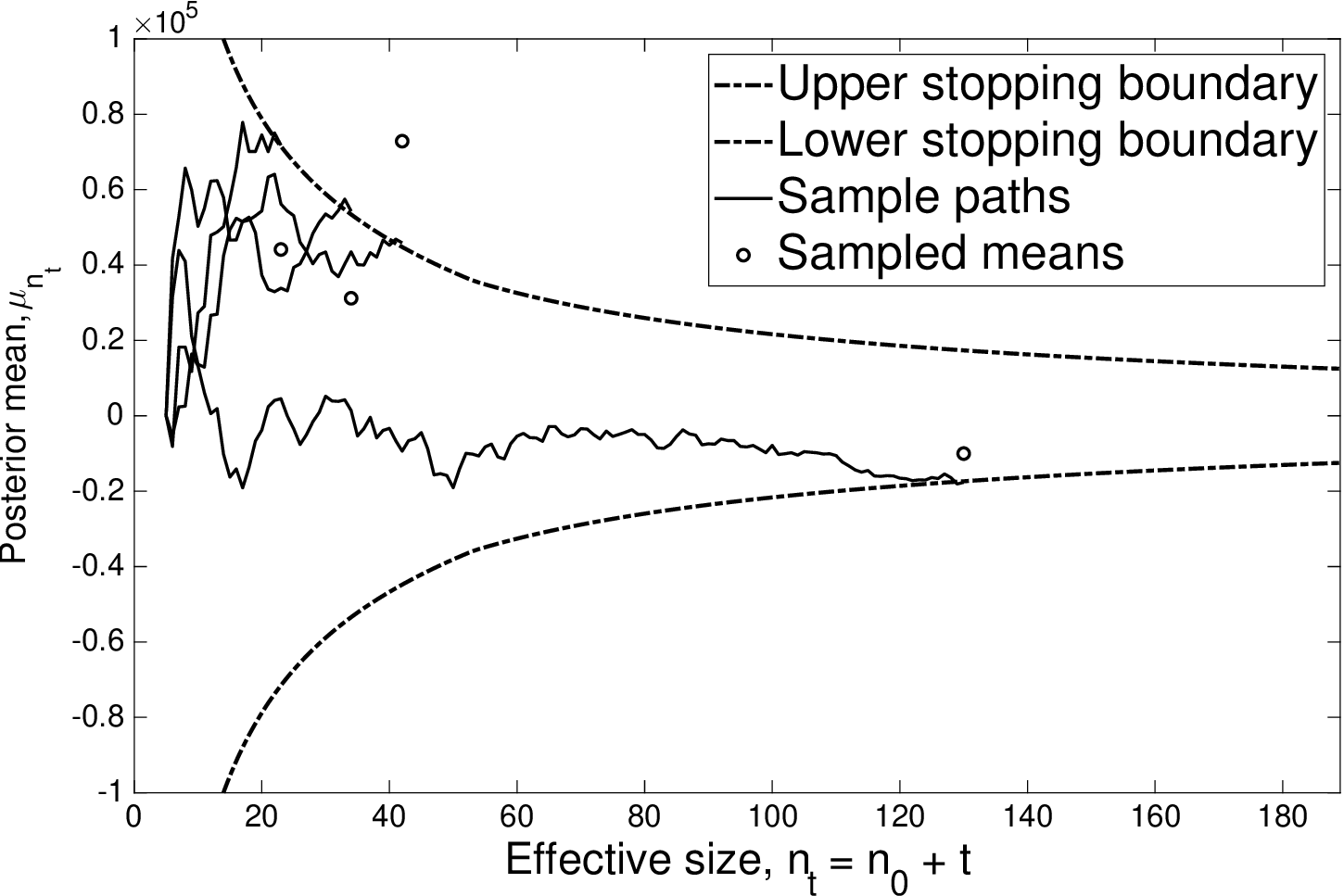}\label{fig:contourYTsamplepaths}}
    \hfill
    \subfloat[Early stopping requires stronger evidence: the posterior probability of incorrect selection (vertical axis) must be smaller to justify stopping with a smaller sample size, $t$.]{\includegraphics[width=0.48\linewidth]{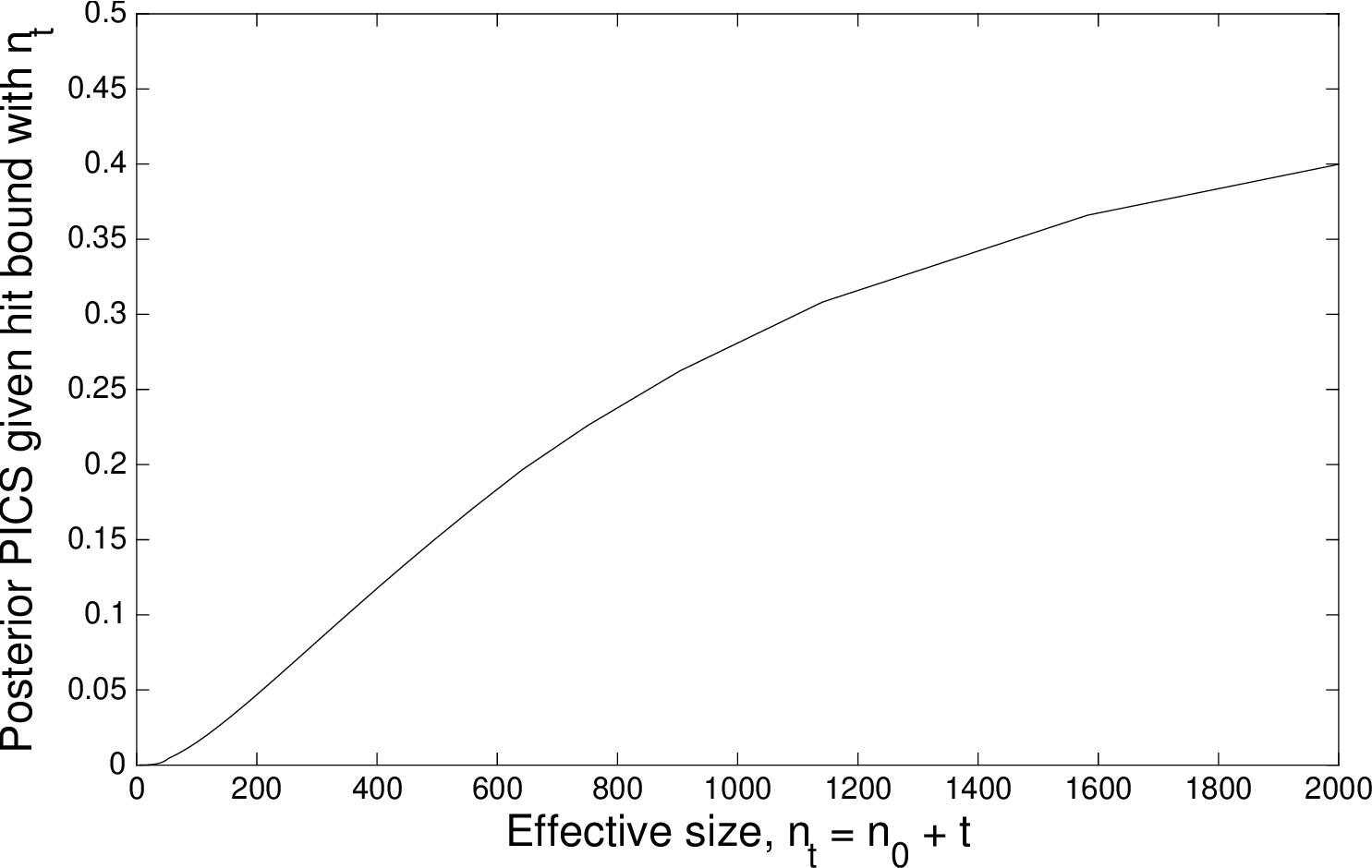}\label{fig:SeqBayesPostPICS}}
    \caption{Fully Sequential economics of selection Procedure ($\sigma=10^5$, $\cost=1$, $
    \mu_0 = 0$, $\nreps{0}=5$).}
\end{figure}

\figref{fig:contourYTsamplepaths} illustrates the optimal stopping boundaries with dashed lines for our running example, with $\cost=1$ and $\sigma=10^5$. It uses the notation $\nreps{t} = \nreps{0} + t$ for the effective number of samples. Four simulated sample paths are drawn. Circles represent the simulated mean for the sample path, which is sampled from the prior distribution. Jagged lines are the sample paths for those realized means, up to the time the sample path first leaves the continuation region (between the stopping boundaries). \textit{This is not a fixed-budget procedure}: there is no one deterministic resource constraint the modeler can specify at time 0. \citet{chick:12a} show that there is a finite upper bound on the stopping time when the sample variance is known. But this does not mean that the procedure is fixed-budget for two reasons. One, the bound emerges from calculations with model parameters, not an inherent organizational budget constraint. Two, the bound does not hold if the variance is unknown, in which case plug-in sample variance estimators can be used  \citep{chick2017bayesian}.

The optimal time $\stoptimeopt$ is when the posterior mean hits the stopping boundary for the first time. Given the symmetry of the continuation region, we have $\myabs{\prmeans{\stoptimeopt}} = \beta^{-1} b(1/(\gamma \nreps{\stoptimeopt}))$, and the posterior variance of the unknown mean is $\sigma^2/\nreps{\stoptimeopt}$. From this, we can compute the posterior probability of incorrect selection, \PICS, $\Phi(- \beta^{-1} b(1/(\gamma \nreps{\stoptimeopt})) / (\sigma/\sqrt{\nreps{\stoptimeopt}}))$, up to a diffusion approximation.
\figref{fig:SeqBayesPostPICS} shows the posterior probability of incorrect selection. The posterior \PICS when optimally stopping with this \EOS \RnS procedure increases with sample size until stopping. This is because losses can be larger with smaller sample sizes, so a greater degree of evidence for correct selection is required with small sample sizes.
\textit{This is not a fixed-precision procedure}: there is no deterministic Bayesian precision target $\alpha$ for \PICS (nor for good selection) that the modeler can specify at time 0.
The posterior \EOC is not constant, either, at stopping (data not shown). Similarly, there are no specific claims of frequentist guarantees for \PICS or \EOC.

We return to our running example, with $\cost = 1,\prmeans{0}=0, \nreps{0}=5$. The expected sample size is $\E[\stoptimeopt|\filtration{0}] = 119.3 \pm 1.6$ (mean $\pm$ stderr from $10^4$ replications), and an expected total cost of 189.0. These Monte Carlo estimates are close to the PDE solution estimates of 119.0 and 194.7, respectively, using publicly available code \citep{ChiForPer:JRSSBcode}.
Thus, both the expected sample size and expected total cost decrease with this fully sequential procedure as compared with the optimal one-stage fixed sample size in \secref{sub:Bayesdetexample} ($207 > 119$ and $418.6 > 194.7$). Thus, response adaptive stopping can significantly cut expected sample size and losses relative to one-stage sampling. However, the variability in sample sizes can be large. Here, $(\Var[\stoptimeopt|\filtration{0}]])^{1/2} = 164.2$.

When observations have unknown variance, it is straightforward to use conjugate normal-inverse-gamma prior distributions for the unknown mean and variance, or to use plug-in estimators for the variance that are updated with the means \citep{chick2015optimal,chick2017bayesian}. Those works also allow for the outcomes to be observed with delay.

\section{Frequentist One-Stage and Two-Stage \EOS Procedures}
\label{sec:freqeos}

When people think of \EOS procedures, Bayesian decision theory comes to mind. Here and in \secref{sec:freqeosseq} we demonstrate that frequentist \EOS procedures can be developed, too.
The Bayesian \EOS approach in \secref{sec:examples} uses a prior distribution over the unknown parameters to average performance over a distribution of configurations.
\RnS procedure performance, conditional at specific configurations, implies a frequentist assessment.

One interpretation of a frequentist \EOS procedure is to assess its performance, conditional on the parameters that define problem configurations, with an indifference-zone framing. The idea is to guarantee the best possible worst-case performance over a set of configurations of interest, in an attempt to guarantee a worst-case expected total cost. We explore this interpretation in this section.

Another interpretation is to abandon the idea of guarantees, noting that guarantees are for constraints on sampling budget or selection quality, and that \EOS has an objective function to minimize, without constraints on sampling costs or selection quality. The idea is to minimize the best approximation for the sum of sampling cost and opportunity costs, in a sequential procedure that updates parameters as data become available, and adapts allocation and stopping decisions based on that accumulating information. We explore this interpretation below in \secref{sec:freqeosseq}.

For the first interpretation, based on indifference-zone ideas, we build step-by-step to two new frequentist \EOS procedures for $\numproj\geq 2$ alternatives with unknown means and variances.
In \secref{sec:freqknownstandard}, we present a one-stage procedure for two alternatives assuming the setup of \secref{sec:examples}: one alternative has a known constant value, the other has an unknown mean and a known sampling variance. This builds on \citet{threeflavorswsc26}.

In \secref{sec:freqeoskbig}, we allow both the means and variances to be unknown in two-stage procedures. Variances are estimated, as is common in fixed-precision IZ procedures, in a first stage of sampling. We show that if the variances are unrestricted, the worst-case expected total cost is unbounded. We then provide two new two-stage \EOS procedures for $\numproj \geq 2$ alternatives with unknown variances with the indifference-zone framing.
The first of these assumes an upper bound on the sample variances. Such an upper bound, when considered in terms of signal-to-noise ratios, is a control on the maximum noise, and is dual to the usual IZ formulation which assumes the mean of the best is $\delta>0$ better than the alternatives, a control on the minimum signal.
The second new two-stage procedure with  $\numproj\geq 2$ alternatives and unknown variances uses the first-stage sampling to provide simultaneous upper confidence bounds constructed from Stage~1 data, and therefore a specified confidence that the total sampling and selection loss has been upper-bounded.

We introduce notation for the problem with $\numproj$ alternatives. Let $\thetavec=(\theta_1,\ldots,\theta_{\numproj})$ be the fixed but unknown vector of means. A policy $\pi$ specifies allocation, stopping, and selection rules. Let $n_i(\pi)$ be the sample size from alternative $i$ when the policy stops, let $c_i$ be the cost per observation, and let $D_\pi$ be the selected alternative. For a fixed configuration $\thetavec$, the expected total cost of $\pi$ is
\begin{equation}
\label{eq:frequentrisk}
R(\pi,\thetavec)
=\underbrace{\E\left[\sum_{i=1}^{\numproj}c_i n_i(\pi) \big| \thetavec \right]}_{
\text{expected sampling cost}}
+\underbrace{\E\left[\max_j\theta_j-\theta_{D_\pi} \big| \thetavec \right]}_{
\text{expected selection loss (EOC)}}.
\end{equation}
The conditional expectations in \eqref{eq:frequentrisk} are over sampling randomness under a fixed configuration $\thetavec$, not an average over a prior distribution on $\thetavec$.

\subsection{One-Stage, Known Variance, $\numproj=2$ with a Known Standard}
\label{sec:freqknownstandard}

We return to the two-alternative model used for the Bayesian procedures in \secref{sec:examples}. Alternative~2 has the known constant value $\theta_2=0$, so there is no value in sampling it. Alternative~1 has fixed but unknown mean $\theta_1=\theta$, and its observations $X_{1j}$ are independent $\Normal(\theta,\sigma^2)$ with known $\sigma^2$ and cost $\cost=c_1>0$ per observation. Consider a one-stage policy $\pi$ that takes a fixed sample size of $n\geq1$ observations from alternative~1. Then $n_1(\pi)=n$ and $n_2(\pi)=0$, so $n$ is the only sample-size decision. Define $\overline X_{1,n}=n^{-1}\sum_{j=1}^nX_{1j}$. The policy selects alternative~1 when $\overline X_{1,n}\geq0$ and selects the constant alternative otherwise; that is, $D_\pi=1$ if $\overline X_{1,n}\geq0$ and $D_\pi=2$ otherwise. Because $\overline X_{1,n}\sim\Normal(\theta,\sigma^2/n)$, its frequentist expected total cost at the fixed mean $\theta$, in \eqref{eq:frequentrisk}, reduces to
\begin{equation}
\label{eq:totalIZcost}
R(n,\theta)
=\cost n+|\theta|\Phi\left(-\frac{|\theta|\sqrt n}{\sigma}\right).
\end{equation}
Indeed, if $\theta>0$, an incorrect selection occurs when $\overline X_{1,n}<0$; if $\theta<0$, it occurs when $\overline X_{1,n}\geq0$. In either case its probability is $\Phi(-|\theta|\sqrt n/\sigma)$ and its opportunity cost is $|\theta|$, which gives \eqref{eq:totalIZcost}.

For each fixed $\theta$, \eqref{eq:totalIZcost} can be minimized over this restricted one-stage sign-selection family. Because $n$ is a positive integer and $R(n,\theta)\geq \cost n\to\infty$, a minimizing sample size exists. We call it an oracle sample size because it depends on $\theta$; the resulting choice is not implementable when the mean is unknown. More strongly, there is generally no single sample size that is optimal for all $\theta$ (see \propref{prop:nooneStagepointwise} in \appref{app:section4aux}).
This lack of a uniformly pointwise-optimal sample size does not arise in the same form for the Bayesian procedure of \citet{degroot:decisions} in \secref{sub:Bayesdetexample}, where averaging $R(n,\theta)$ over the prior distribution of $\theta$ produces a single prior expected loss that can be minimized without knowing the realized value of $\theta$.
For a frequentist benchmark, we instead choose the sample size that minimizes the highest expected total cost over the specified set of means. Taking a supremum over fixed configurations parallels the uniform protection used in traditional fixed-precision \RnS.

Let $\Theta_\delta=\{\theta:|\theta|\geq\delta\}$, where $\delta\geq0$ is typically referred to as the indifference-zone (IZ) parameter. When $\delta>0$, configurations with $|\theta|<\delta$ are excluded; when $\delta=0$, all $\theta\in\mathbb R$ are included. We consider the one-stage minimax \EOS problem
\begin{equation}
\label{def:econselectIZ}
\min_{n>0}J_\delta(n),
\qquad
J_\delta(n)
:=\sup_{\theta\in\Theta_\delta}R(n,\theta)
=\cost n+\sup_{|\theta|\geq\delta}
|\theta|\Phi\left(-\frac{|\theta|\sqrt n}{\sigma}\right).
\end{equation}
Here $n>0$ is continuous. Because \(J_\delta(n)\to\infty\) as \(n\downarrow0\) and as \(n\to\infty\), $J_\delta(n)$ has a finite positive minimizer. We first solve this relaxation and then recover the integer optimum.

We solve the inner maximization before optimizing the sample size. For fixed $n$ and $\theta$, the selection loss is the second term of $R(n,\theta)$: the opportunity cost $|\theta|$ multiplied by the probability of incorrect selection $\Phi(-|\theta|\sqrt n/\sigma)$. Let $z=|\theta|\sqrt n/\sigma$ be the standardized mean gap and define $f(z)=z\Phi(-z)$ for $z\geq0$. Because $|\theta|=\sigma z/\sqrt n$, the restriction $|\theta|\geq\delta$ is equivalent to $z\geq\delta\sqrt n/\sigma$, and the inner supremum in \eqref{def:econselectIZ} becomes $\frac{\sigma}{\sqrt n} \sup_{z\geq\delta\sqrt n/\sigma}f(z)$. Thus the inner problem is precisely to maximize $f$ over the interval $[\delta\sqrt n/\sigma,\infty)$. By \lemref{lem:normaltailshape} in \appref{app:section4aux}, $f$ is strictly increasing and then strictly decreasing, with unique maximizer $z^*>0$ characterized as the unique positive root of
$\Phi(-z)=z\phi(z)$.
Thus $z^*$ is the standardized gap that produces the largest normalized selection loss. Define $f^*=f(z^*)$. Numerically, $z^*=0.7517915, \Phi(-z^*)=0.2260882$, and $f^*=0.1699712$.

We can then derive the least-favorable configurations (LFCs) of $\theta$ that attain the supremum in \eqref{def:econselectIZ} for each fixed $n$; equivalently, these are the values of $\theta$ that maximize the selection loss.
\begin{lemma}[Least-favorable configurations]
\label{lem:IZk2}
Fix $n>0$. The LFCs for the supremum in \eqref{def:econselectIZ} are
\begin{equation}
\label{eq:thetaLFC}
\theta_{\rm LFC}(n)\in
\left\{
-\max\left\{\delta,\frac{z^*\sigma}{\sqrt n}\right\},
\ \max\left\{\delta,\frac{z^*\sigma}{\sqrt n}\right\}
\right\}.
\end{equation}
Thus, with $n^+_\delta=(z^*\sigma/\delta)^2$ for $\delta>0$ and $n^+_\delta=\infty$ for $\delta=0$, the worst-case expected total cost is
\begin{equation}
\label{eq:freqriskpiecewise}
J_\delta(n)
=\sup_{\theta\in\Theta_\delta}R(n,\theta)
=\begin{cases}
\cost n+\sigma f^*/\sqrt n,
&0<n\leq n^+_\delta,\\[4pt]
\cost n+\delta\Phi(-\delta\sqrt n/\sigma),
&n\geq n^+_\delta.
\end{cases}
\end{equation}
\end{lemma}

A proof is provided in \appref{app:section4aux}. The IZ-free setting corresponds to $\delta=0$. Then $n^+_\delta=\infty$, so only the first case occurs for every $n>0$: $\theta_{\rm LFC}(n)=\pm z^*\sigma/\sqrt n$ and $\PICS_{\rm LFC}(n)=\Phi(-z^*)=0.2261$. Thus, the procedure can be IZ-free while still yielding a finite sample size, as with the procedure of \citet{FanHongNelson2016izfree}.

Having evaluated the supremum over $\theta$ for every $n$, we now minimize the explicit function $J_\delta(n)$ in \eqref{eq:freqriskpiecewise} for a given $
\delta
\geq 0$. Define the sampling-cost threshold
\begin{equation*}
\cost_\delta=\frac{\delta^3\phi(z^*)}{2\sigma^2z^*}
=\frac{f^*\delta^3}{2(z^*)^3\sigma^2}.
\end{equation*}
For $\delta>0$, differentiating either piece of \eqref{eq:freqriskpiecewise} at $n^+_\delta$ gives $ J_\delta'(n^+_\delta)=\cost-\cost_\delta. $ Thus $\cost_\delta$ is the marginal reduction in worst-case selection loss at $n=n^+_\delta$, where $z^*\sigma/\sqrt{n^+_\delta}=\delta$.

\begin{theorem}[Minimax \EOS sample size]
\label{thm:freqoptimaln}
The function $J_\delta$ in \eqref{eq:freqriskpiecewise} is continuously differentiable and strictly convex. Its unique continuous minimizer is
\begin{equation}
\label{eq:freqoptimaln}
n^*=
\begin{cases}
\left(\dfrac{\sigma f^*}{2\cost}\right)^{2/3},
&\cost\geq\cost_\delta,\\[10pt]
\dfrac{\sigma^2}{\delta^2}
W_0\left(\dfrac{\delta^6}{8\pi\cost^2\sigma^4}\right),
&0<\cost<\cost_\delta,
\end{cases}
\end{equation}
where $W_0(x)$ is the principal branch of the Lambert $W$ function: for $x\geq0$, $W_0(x)$ is the unique nonnegative value $w$ satisfying $we^w=x$. Because the one-stage policy requires $n\geq1$, let $\bar n^*=\max\{1,n^*\}$. The integer optimum is obtained by comparing $J_\delta(\lfloor\bar n^*\rfloor)$ and $J_\delta(\lceil\bar n^*\rceil)$.
\end{theorem}

This result, proven in \appref{app:section4aux}, allows us to define \textbf{Procedure \EOSonekone} (subscript 1 for one-stage) for comparing $\numproj=1$ unknown mean with 0, given parameters $\sigma>0, \cost>0, \delta\geq 0$. First, observe outcomes with the optimal integer sample size $n^*$, as discussed in \thmref{thm:freqoptimaln}, then choose the alternative with unknown mean as best if and only if the sample mean exceeds 0.

The following consequences of \thmref{thm:freqoptimaln} and \lemref{lem:IZk2} describe the unconstrained continuous minimizer $n^*$. If $\cost\geq\cost_\delta$, then $n^*\leq n^+_\delta$. Therefore, as in the discussion following \lemref{lem:IZk2}, the two LFC means are $\theta_{\rm LFC}(n^*)=\pm z^*\sigma/\sqrt{n^*}$, and $\PICS_{\rm LFC}(n^*)=\Phi(-z^*)=0.2261$.
When $\delta=0$, $\cost_\delta=0$ and $n^+_\delta=\infty$, so every $\cost>0$ falls in this first case.

If $0<\cost<\cost_\delta$, then $n^*>n^+_\delta$. The two LFC means are therefore $\theta=-\delta$ and $\theta=\delta$, and $\PICS_{\rm LFC}(n^*)=\Phi(-\delta\sqrt{n^*}/\sigma)$. Substituting the second formula for $n^*$ in \thmref{thm:freqoptimaln} gives $\PICS_{\rm LFC}(n^*)= \Phi\!\left(-\sqrt{W_0(\delta^6/(8\pi\cost^2\sigma^4))}\right)$.
Within this case, $\PICS_{\rm LFC}(n^*)$ is increasing in $\cost$. As $\cost\downarrow0$, $n^*\to\infty$ and $\PICS_{\rm LFC}(n^*)\to0$, or equivalently $\PCS_{\rm LFC}(n^*)\to1$. Thus, \emph{cheaper sampling leads to a larger optimal sample size and improves selection quality at the LFCs.}

\begin{figure}[tb]
    \centering
    \subfloat[Optimal sample size, $n^\ast $.] {
        \includegraphics[width=0.31\linewidth]{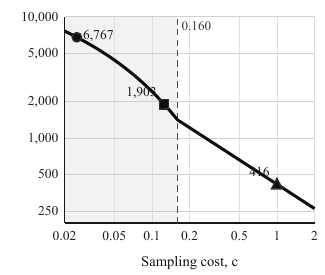}
        \label{fig:freqeconomicnstar}}
    \subfloat[PICS at the LFC, $\theta_{\rm LFC}(n^\ast )$.] {
        \includegraphics[width=0.31\linewidth]{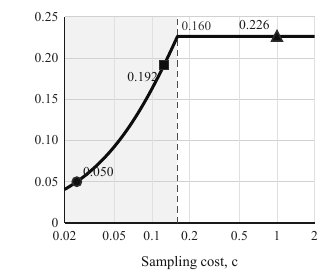}
        \label{fig:freqeconomicpics}}
    \subfloat[Best worst-case expected total cost, $J_\delta(n^\ast )$.] {
        \includegraphics[width=0.35\linewidth]{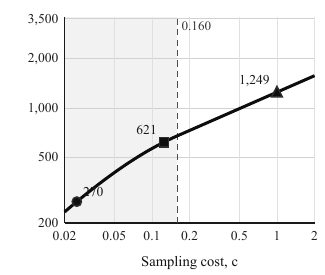}
        \label{fig:freqeconomictotalcost}}
    \caption{Performance of Procedure \EOSonekone in the running example with
    $\sigma=10^5$ and $\delta=2,000$. The vertical dashed line is at
    $\cost_\delta=0.16$. The circle, square, and triangle mark
    $\cost=1/39.9$, $1/8$, and $1$, respectively. Both axes in (a) and (c),
    and the horizontal axis in (b), are logarithmic.
    \label{fig:freqeconomic}}
\end{figure}

For the running example, $\sigma=10^5$, $\delta=2,000$, and $\cost=1$ give $\cost_\delta=0.1600$ and $n^*=416.45$.
\figref{fig:freqeconomic} shows the sample size that optimizes expected total cost, the PICS at the LFCs, and the minimized worst-case expected total cost as functions of the sampling cost $\cost$.

\begin{remark}
\label{rem:twostochastic}
The comparison with a known standard extends directly to two alternatives with unknown means \(\theta_1\) and \(\theta_2\), known positive sampling variances \(\sigma_i^2\), and sampling costs \(c_i\). The optimal (continuous) allocation with $\numproj=2$ equates the marginal reduction in the variance of the difference of the two sample means ${\sigma_1^2}/{n_1}+{\sigma_2^2}/{n_2}$ per unit sampling cost, so \(\sigma_1^2/(c_1 n_1^2)=\sigma_2^2/(c_2 n_2^2)\) and the selection problem then becomes a comparison of the unknown mean difference \(\theta_1-\theta_2\) with the known standard zero: select alternative~1 when \(\overline X_{1,n_1}-\overline X_{2,n_2}\geq0\), and alternative~2 otherwise.
\end{remark}

\subsection{Two-Stage, Multiple Alternatives with Potentially Unknown Variances}
\label{sec:freqeoskbig}

We now extend the one-stage global worst-case formulation to a setting with $\numproj\geq 2$ alternatives, in which the means are unknown, and the variances may be unknown and differ.  Observations $
X_{i,j}$ from alternative $i$ cost $\costs{i}$ each and are distributed $\Normal(\theta_i,\sigma_i^2)$, jointly independent across $i$ and $j$. Write $\vecsigsq=(\sigma_1^2,\sigma_2^2,\ldots,\sigma_{\numproj}^2)$ and $\costvec = (\costs{1},\costs{2},\ldots,\costs{\numproj})$.

We first explore the class of two-stage procedures, where a first stage of sampling estimates unknown variances and a second stage of sampling gives additional information for selection. We first state the global worst-case criterion and identify a challenge if the unknown variances do not have a known upper bound. We then propose two approaches to address that challenge. A frequentist \EOS procedure for multiple alternatives with known variances emerges as a corollary.

Let $\Pi$ be a candidate class of sampling-and-selection policies. For $\pi\in\Pi$, let $N_i(\pi)$ be its total sample size from alternative $i$ and let $D_\pi$ be its selected alternative. To make explicit the variance dependence that is suppressed in the general notation of \eqref{eq:frequentrisk}, define
\begin{equation}
\label{eq:unknownvarrisk}
R(\pi,\thetavec,\vecsigsq)
=\underbrace{\sum_{i=1}^{\numproj}c_i\E[N_i(\pi)]}_{
\text{expected sampling cost}}
+\underbrace{\E\left[
\max_{1\leq j\leq\numproj}\theta_j-\theta_{D_\pi}
\right]}_{\text{expected selection loss (EOC)}}.
\end{equation}
We seek policies that perform well under the global worst-case formulation in \eqref{eq:unknownvarglobalproblem}.
\begin{equation}
\label{eq:unknownvarglobalproblem}
\inf_{\pi\in\Pi}\ \sup_{\substack{
\thetavec\in\mathbb R^{\numproj}\\
\vecsigsq\in(0,\infty)^{\numproj}}}
R(\pi,\thetavec,\vecsigsq).
\end{equation}

We build on one of the two-stage procedures of \citet{chick:izopcost} that delivers frequentist fixed-precision $\EOC \leq \EOCmax$ guarantees, for a user-specified $
\EOC$ bound, $\EOCmax$.

\newcommand{\proceoc}[2]{OC$_f(#1,#2)$}
\smallskip
\noindent\textbf{Procedure \proceoc{g}{n_0}:} One can specify either $g$ or $\EOCmax$ to determine the other.
\begin{enumerate}
\item \textbf{Setup.} For user-specified upper bound $
\EOCmax$ on $\EOC$, use a table or computation from \citet{chick:izopcost} to find the constant $\zeta_{\numproj,n_0}$, and compute $g=\zeta_{\numproj,n_0} / \EOCmax$.
\item \textbf{Stage 1.} For each alternative $i=1,\ldots,\numproj$, collect $n_0\geq3$ independent observations and compute the unbiased sample variance $S_i^2$.
\item \textbf{Allocation.} Set the total sample size for alternative $i$ to
\begin{equation}
\label{eq:twostagesamples}
N_i(g)=\max\left\{n_0,\left\lceil g^2 S_i^2\right\rceil\right\},
\qquad i=1,\ldots,\numproj,
\end{equation}
and collect $N_i(g)-n_0$ additional observations from alternative $i$.
\item \textbf{Selection.} Using all $N_i(g)$ observations, compute the overall sample mean $\overline X_i(N_i(g))=N_i(g)^{-1}\sum_{r=1}^{N_i(g)}X_{i,r}$ and select an alternative in $\arg\max_{1\leq i\leq\numproj}\overline X_i(N_i(g))$, breaking ties arbitrarily.
\end{enumerate}
\smallskip

Denote this policy by $\pi_g$, so $N_i(\pi_g)=N_i(g)$ and $D_{\pi_g}$ is the alternative selected in Step~3. Thus $N_i(g)$ includes the $n_0$ Stage~1 observations, and a larger $g$ generally produces larger final sample sizes.
\cite{chick:izopcost} show that, for each $\numproj$ and $n_0$, a constant $\zeta_{\numproj,n_0}>0$ can be computed before sampling such that Procedure~\proceoc{g}{n_0} satisfies the EOC bound.
We rewrite that bound in a form useful for analysis below, and set  $\zeta=\zeta_{\numproj,n_0}$.
\begin{equation}
\label{eq:eocscaleguarantee}
\sup_{\substack{
\thetavec\in\mathbb R^{\numproj}\\
\vecsigsq\in(0,\infty)^{\numproj}}}
\E\left[\max_j\theta_j-\theta_{D_{\pi_g}}\right]
\leq\frac{\zeta}{g}.
\end{equation}
We observe that increasing $g$ reduces the EOC bound $\zeta/g$ but raises the expected sampling cost.

To analyze the expected sampling cost, we derive the expected sample size produced by the procedure. Let $F_\nu$ and $\overline F_\nu=1-F_\nu$ denote the CDF and survival function of a chi-squared random variable $\chi_\nu^2$ with $\nu$ degrees of freedom. Let $\nu=n_0-1$, so that $\nu S_i^2/\sigma_i^2\sim\chi_\nu^2$.

We define the continuous approximation to the sample size,
\[
\widetilde N_i(g)=\max\{n_0,g^2S_i^2\}.
\]

\begin{proposition}[Expected sample size]
\label{prop:expectedtwostagesamples}
For every $g>0$ and $\sigma_i^2>0$, define $a_i(g)=\frac{\nu n_0}{g^2\sigma_i^2}$, then
\begin{eqnarray}
    \label{eq:exactexpectedN}
\E[\widetilde N_i(g)] &=&
    n_0F_\nu(a_i(g))+g^2\sigma_i^2\overline F_{\nu+2}(a_i(g)) \mbox{, and } \\
\label{eq:exactexpectedNderivative}
\frac{d}{dg}\E[\widetilde N_i(g)]&=&
    2g\sigma_i^2\overline F_{\nu+2}(a_i(g))>0.
\end{eqnarray}
Moreover, the expected sample size is strictly increasing in the sampling variance:
\begin{equation}
\label{eq:exactexpectedNvariancederivative}
\frac{\partial}{\partial\sigma_i^2}\E[\widetilde N_i(g)]
=g^2\overline F_{\nu+2}(a_i(g))>0.
\end{equation}
The error from using this continuous approximation satisfies
$0\leq\E[N_i(g)]-\E[\widetilde N_i(g)]<1.
$\end{proposition}

See \appref{app:section5proofs} for proof. Thus \eqref{eq:exactexpectedN} gives the expected final sample size up to less than one observation per alternative.
Combining the selection-loss bound in \eqref{eq:eocscaleguarantee} with the expected-sample-size bound in Proposition~\ref{prop:expectedtwostagesamples} gives, for every fixed $g>0$, Procedure~\proceoc{g}{n_0} satisfies
\begin{equation}
\label{eq:unrestrictedCWbound}
\frac{\zeta}{g} + \sum_{i=1}^{\numproj}c_i\left[
\sup_{\sigma_i^2>0}\E[\widetilde N_i(g)] - n_0
\right]
\leq
\sup_{\substack{
\thetavec\in\mathbb R^{\numproj}\\
\vecsigsq\in(0,\infty)^{\numproj}}}
R(\pi_g,\thetavec,\vecsigsq)
\leq
\frac{\zeta}{g} + \sum_{i=1}^{\numproj}c_i\left[
\sup_{\sigma_i^2>0}\E[\widetilde N_i(g)]+1
\right].
\end{equation}
The expectations in \eqref{eq:unrestrictedCWbound} are given explicitly by \eqref{eq:exactexpectedN}, the $+1$ in the second expectation allows for rounding the sample size, and the $-n_0$ in the first expectation accounts for the Stage~1 samples in the $\min$ that defines the $N_i$.
Thus, within the Chick--Wu family, the right-hand side of \eqref{eq:unrestrictedCWbound} is a good upper bound for the \EOS in \eqref{eq:unknownvarglobalproblem}. \emph{We will optimize $g$ to obtain two-stage \EOS procedures}.

\propref{prop:frequnbounded} states that a global worst case over unbounded variances is infinitely conservative for the expected total cost.
Its proof is found in \appref{app:section5proofs}.

\begin{proposition}[Unbounded worst-case total cost]
\label{prop:frequnbounded}
For every fixed $g>0$ and every alternative $i$,
\[
\sup_{\sigma_i^2>0}\E[N_i(g)]=
\sup_{\sigma_i^2>0}\E[\widetilde N_i(g)]=\infty.
\]
\end{proposition}

\begin{remark}
\propref{prop:frequnbounded} poses a challenge. If the set of configurations of interest allows at least one alternative to have unbounded variance, there is no well-defined \EOS procedure that minimizes the bounds on worst-case expected total cost in \eqref{eq:unrestrictedCWbound}, because those bounds would be unbounded.
\end{remark}

We therefore consider and analyze two scenarios. One, the decision maker may have deterministic upper bounds $\overline\sigma_i^2$ on the unknown sampling variances. An important special case is that the variances are known to be exactly $\overline\sigma_i^2$. Two, if such information is unavailable, Stage~1 data can be used to construct simultaneous upper confidence bounds for the variances.

\subsubsection{Two-Stage Frequentist \EOS: Known Upper Bounds on Variances}
\label{sec:freqgeneraleos}

Suppose that there are known positive bounds (or ``caps'') $\overline\sigma_i^2$ for the potentially unknown variances for all alternatives.  Write $\vecsigmabnd =
(\overline\sigma_1,\ldots,\overline\sigma_{\numproj})$
and define the variance-capped configuration class
\begin{equation*}
\Xi_{\vecsigmabnd}
=\left\{(\thetavec,\vecsigsq):
\thetavec\in\mathbb R^{\numproj},\
0<\sigma_i^2\leq\overline\sigma_i^2<\infty,
\ i=1,\ldots,\numproj\right\}.
\end{equation*}
By \eqref{eq:exactexpectedNvariancederivative}, $\E[\widetilde N_i(g)]$ is strictly increasing in $\sigma_i^2$. Therefore, when the unrestricted supremum in \eqref{eq:unrestrictedCWbound} is replaced by $0<\sigma_i^2\leq\overline\sigma_i^2$, the worst-case expected sample size is attained at $\sigma_i^2=\overline\sigma_i^2$. Define
\[
\overline a_i(g)=\frac{\nu n_0}{g^2\overline\sigma_i^2}.
\]
Consider the worst-case losses when the set of configurations of interest is the variance-capped class, $\Xi_{\vecsigmabnd}$, which is bounded above by:
\begin{equation}
\label{eq:tighterriskbound}
\sup_{(\thetavec,\vecsigsq)
\in\Xi_{\vecsigmabnd}}
R(\pi_g,\thetavec,\vecsigsq)
\leq\sum_{i=1}^{\numproj}c_i\left[
n_0F_\nu(\overline a_i(g))
+g^2\overline\sigma_i^2\overline F_{\nu+2}(\overline a_i(g))
\right]+\frac{\zeta}{g}+\sum_{i=1}^{\numproj}c_i.
\end{equation}
The upper bound in \eqref{eq:tighterriskbound} contains the exact continuous expected sample sizes at the standard deviation caps, while $\sum_i c_i$ allows for rounding the sample size. We can then optimize $g$.

\begin{proposition}[Sample size scaling parameter, $g$, with bounded variances]
\label{prop:unknownvarchirefinement}
The upper bound in \eqref{eq:tighterriskbound} is strictly convex in $g$ and has a unique minimizer $\widetilde g^*>0$, characterized by
\begin{equation}
\label{eq:tightergFOC}
2(\widetilde g^*)^3\sum_{i=1}^{\numproj}c_i\overline\sigma_i^2
\overline F_{\nu+2}\left(\frac{\nu n_0}{(\widetilde g^*)^2\overline\sigma_i^2}\right)
=\zeta.
\end{equation}
\end{proposition}

A proof is provided in \appref{app:section5proofs}.
The unique solution of \eqref{eq:tightergFOC} can be found by one-dimensional root finding.

\begin{corollary}
\label{cor:boundedvar}
Procedure \proceoc{\widetilde g^*}{n_0} with $\EOC$ bound $\EOCmax=\zeta_{\numproj,n_0} / \widetilde g^*$ is a new two-stage frequentist \EOS procedure that we call \textbf{Procedure \twoseosvarbound},
for the variance-capped configuration class $\Xi_{\vecsigmabnd}$.
The LFCs for this class have the variances equal to their specified upper bounds.
\end{corollary}

An upper bound on the variance of outcomes very naturally complements the widely-studied IZ framework. An IZ bound $\delta > 0$ prevents Stage~2 sample sizes from going to infinity in many procedures, and is interpretable as a minimal difference worth detecting. The upper bound on variances also prevents Stage~2 sample sizes from going to infinity in analogous procedures, and is interpretable as the inverse of the minimum (Fisher) information per observation. Both approaches constrain the signal-to-noise ratio of observations in some way.

\begin{corollary}
\label{cor:knownvar}
 \textbf{Procedure \twoseosvarbound}
is also a frequentist \EOS procedure if variances are known, $\sigma^2_i = \bar\sigma^2_i$, for $i=1,2,\ldots,\numproj$.
\end{corollary}

\subsubsection{What if bounds on the variances are not known?}
\label{sec:freqgeneraleosunkvar}

\secref{sec:freqgeneraleos} gave results for $\numproj\geq 2$ when the variances may be unknown but have known upper bounds. In practice, we might not know the variances, nor have good bounds on them. We propose two different procedures to handle unknown variances with two-stage \EOS procedures.

The first, \textbf{Procedure \twoseosvarplugin}, is a naive heuristic that uses the sample variance, $\Svecsq$, for the (unknown) true variances.
It is the same as Procedure \twoseosvarbound from \secref{sec:freqgeneraleos}, \textit{except} that the configuration set of interest is $\Xi_{\vecsigmabnd}$, chosen with heuristic plug-in substitution $\vecsigsqbnd \leftarrow \Svecsq$.

A second approach is to use $n_0$ Stage~1 observations from each alternative to estimate simultaneous $1-\alphavar$ upper confidence bounds on the unknown variances, and define a \textit{conditionally} worst-case  expected total cost over the resulting simultaneous confidence region of the unknown variances, given Stage~1 data. Those upper confidence bounds are (conditionally) assumed to be upper bounds on the variances. We implemented this, splitting the $\alphavar$ error over the $\numproj$ variance estimates, to obtain a \textbf{Procedure \twoseosvarconfidence}. This tends to sample more, relative to \twoseosvarbound, giving a smaller selection loss, and for examples below, the expected total cost of \twoseosvarconfidence was higher than that of \twoseosvarbound. As a result, we do not give our detailed derivation here, and reserve this as an area for potential further research.

\section{New Sequential Frequentist Economics of Selection (\EOS) Procedure}
\label{sec:freqeosseq}

The preceding procedures choose their sample sizes in advance or after one initial stage, with a goal of obtaining a good worst-case bound on expected total cost over problem configurations. As noted in the introduction to \secref{sec:freqeos}, another interpretation of frequentist \EOS is to abandon the worst-case IZ-type bound, and instead focus on minimizing the total sampling cost plus selection loss without worrying about guarantees. With this mindset, a fully sequential \EOS procedure must repeatedly answer two questions: which alternative should receive the next observation, and is the expected benefit of any further observation larger than its cost? The fixed-budget perspective suggests comparing marginal reductions in \EOC with marginal sampling costs. We make that comparison first in an oracle calculation and then replace its unknown quantities by sample estimates.  This results in a new, fully sequential frequentist \EOS procedure.

To derive the marginal reduction in \EOC, fix independent observations $X_{ij}\sim\Normal(\theta_i,\sigma_i^2)$ and evaluate a deterministic allocation $\nvec=(n_1,\ldots,n_{\numproj})$ at a fixed parameter configuration. The calculation is an oracle calculation because it uses the unknown means and variances; it is not an implementable policy. Suppose the best alternative $b=\argmax_i\theta_i$ is unique, let $\Delta_i=\theta_b-\theta_i$ for $i\neq b$, and select the alternative with the largest sample mean. The exact \EOC is
\begin{equation}
\label{eq:sequentialexacteoc}
L(\nvec)
=\sum_{i\neq b}\Delta_i
\Pr\{\overline X_i=\max_j\overline X_j\},
\qquad
\Pr\{\overline X_i=\max_j\overline X_j\}
=\int_{-\infty}^{\infty}p_i(x)\prod_{j\neq i}F_j(x)\,dx,
\end{equation}
where $p_i$ and $F_i$ are the density and CDF of the sample mean $\overline X_i\sim\Normal(\theta_i,\sigma_i^2/n_i)$. The integral permits numerical evaluation of $L(\nvec)$ and hence numerical optimization of $L(\nvec)+\sum_i c_in_i$---or of $L(\nvec)$ subject to a sample-size constraint---for moderate $\numproj$.

A simpler upper bound is available. The variance of the sample-mean difference $\overline X_b-\overline X_i$ is
\begin{equation*}
V_i(\nvec)=\frac{\sigma_b^2}{n_b}+\frac{\sigma_i^2}{n_i},
\end{equation*}
We observe that, for each $i\neq b$,
\[
\Pr\{\overline X_i=\max_j\overline X_j\} \leq\Pr\{\overline X_i>\overline X_b\} =\Phi\left(-{\Delta_i}\big/{\sqrt{V_i(\nvec)}}\right).
\]
Multiplying this inequality by $\Delta_i$ and summing over $i\neq b$ gives
\begin{equation}
\label{eq:sequentialpairbound}
L(\nvec)
\leq\widetilde L(\nvec)
:=\sum_{i\neq b}\Delta_i
\Phi\left(-{\Delta_i}\big/{\sqrt{V_i(\nvec)}}\right).
\end{equation}
We temporarily allow each $n_i$ to vary continuously. Then $M_i(\nvec)=-\partial\widetilde L/\partial n_i$ is the local reduction in the \EOC bound from increasing $n_i$, in the same units as the observation cost $c_i$. Differentiating,
\begin{align}
\label{eq:sequentialmarginalnonbest}
M_i(\nvec)
&=\frac{\Delta_i^2\sigma_i^2}
{2n_i^2V_i(\nvec)^{3/2}}
\phi\left({\Delta_i}\big/{\sqrt{V_i(\nvec)}}\right),
&&i\neq b,\\
\label{eq:sequentialmarginalbest}
M_b(\nvec)
&=\sum_{i\neq b}
\frac{\Delta_i^2\sigma_b^2}
{2n_b^2V_i(\nvec)^{3/2}}
\phi\left({\Delta_i}\big/{\sqrt{V_i(\nvec)}}\right).
\end{align}
Thus, the derivative of the oracle's surrogate objective
$$
\widetilde L(\nvec)+\sum_i c_in_i
$$
with respect to $n_i$ is $c_i-M_i(\nvec)$.
Importantly, for a sequential stopping rule, increasing $n_i$ locally lowers this surrogate objective if and only if $M_i>c_i$.

To obtain a calculable allocation rule and stopping rule, begin Stage~1 sampling with $n_0\geq2$ observations from every alternative, then begin a sequential stage of sampling. Before each potential new observation, let $n_i$, $\overline X_i$, and $S_i^2$ denote the current sample size, sample mean, and sample variance. Let $\widehat b=\argmax_i\overline X_i$ and set $\widehat\Delta_i=\overline X_{\widehat b}-\overline X_i$. Define
\begin{equation*}
\widehat V_i
=\frac{S_{\widehat b}^2}{n_{\widehat b}}
+\frac{S_i^2}{n_i},
\qquad i\neq\widehat b.
\end{equation*}
Substitution into \eqref{eq:sequentialmarginalnonbest}--\eqref{eq:sequentialmarginalbest} gives
\begin{align}
\label{eq:sequentialestimatednonbest}
\widehat M_i
&=\frac{\widehat\Delta_i^2S_i^2}
{2n_i^2\widehat V_i^{3/2}}
\phi\left({\widehat\Delta_i}\big/{\sqrt{\widehat V_i}}\right),
&&i\neq\widehat b,\\
\label{eq:sequentialestimatedbest}
\widehat M_{\widehat b}
&=\sum_{i\neq\widehat b}
\frac{\widehat\Delta_i^2S_{\widehat b}^2}
{2n_{\widehat b}^2\widehat V_i^{3/2}}
\phi\left({\widehat\Delta_i}\big/{\sqrt{\widehat V_i}}\right).
\end{align}
With this continuous local approximation, $c_i-\widehat M_i$ is the estimated change in the surrogate objective per additional observation from alternative $i$.

This defines the first fully sequential, frequentist \EOS procedure,  \textbf{Procedure \seqeos}:
\begin{enumerate}
\item \textbf{Initialization} Take $n_0 \geq 2$ observations from each alternative.
\item \textbf{Update} Update the sample statistics estimates of sample means $\overline X_i$, current best $\hat{b}$, sample variances $S^2_i$, and the terms $\widehat\Delta_i, \widehat{V}_i, , \widehat{M}_i$.
\item \textbf{Stopping rule} If $\max_i\{\widehat M_i-c_i\}\leq0$, Then Stop - go to Selection rule; Else: continue with Allocation rule.
\item \textbf{Allocation rule.} Take one observation from any alternative attaining $\max_i\{\widehat M_i-c_i\}>0$, then continue by going to Update.
\item \textbf{Selection rule.} Pick the alternative with the largest overall sample mean as best.
\end{enumerate}

For common costs $c_i = c$, this means sampling an alternative with the largest $\widehat M_i$ and stopping when $\max_i\widehat M_i\leq\cost$. With heterogeneous costs, the difference $\widehat M_i-c_i$ approximates the change in the expected total cost per additional observation.

We identify some (but certainly not all!) links to some related past work.
One, the term $\widehat{V}_i$ is reminiscent of the complete expected improvement (CEI) of \citet{saleminelsonstaum14CEI}, which has attractive asymptotic fixed-budget properties \citep{ryzhov19cei}.
Two, a Slepian-type inequality rather than a Bonferroni-type inequality \citep{Kim2006} in \eqref{eq:sequentialpairbound} may give a tighter \EOC bound.
Three, the use of plug-ins for sample means and variances is similar to plug-ins for \OCBA-type fixed-precision procedures \citep[\S3.4]{chenlee2011ocbabook}, though our plug-ins assess the objective function rather than selection quality constraints.
Four, there may be value in using the rate of improvement per unit sampling budget, $\widehat M_i/c_i$, for allocation and stopping rules and multi-step lookaheads, as with $KG^*$ indices \citep{chickfrazier:wsc09}, or even multi-step for multi-arm lookaheads, rather than the rate of improvement per observation.
Five, performance might improve with bias corrections for $\widehat{M}_i$, a nonlinear function of sample means, using second-order Taylor series expansions or jackknife estimators \citep{bratley:simulation}.
These represent areas for future work.

\section{Numerical Study}
\label{sec:numericalstudy}

We explore several questions about the performance of the new two-stage and fully sequential \EOS procedures proposed above with simulations. What is the penalty of not accurately and tightly specifying the upper bound on the standard deviations that are used in the two-stage procedures?
How do the several two-stage \EOS procedures compare in performance, given their different ways of handling uncertainty about the variances of the observations?
How do the two-stage \EOS procedures compare with two-stage fixed-budget and fixed-precision procedures?
What are some strengths and weaknesses of the mechanism proposed above for the first fully sequential frequentist \EOS procedure, relative to two-stage procedures?
The subsections below assess these questions with parameters chosen to illustrate the main concepts.

Before presenting this analysis, we describe the experimental testbed. We describe the procedures and comparator procedures, which include procedures that are two-stage and sequential, frequentist and Bayesian, as well as fixed-budget and fixed-precision.
We then describe how we set parameters of the procedures, and performance measures used to compare the procedures.
We also describe our choice of problem configurations used to demonstrate the procedures, and other specific implementation details for the analysis.

\paragraph{Procedures and comparators.}
We compare our new multi-stage frequentist \EOS procedures (\twoseosvarbound, \twoseosvarplugin, and \twoseosvarconfidence from \secref{sec:freqeos}, which build on the Chick-Wu two-stage procedure and make different assumptions about the sampling variances, and the new sequential \seqeos procedure in \secref{sec:freqeosseq}) with several comparators.
The comparators are two fixed precision procedures (the two-stage procedure of \citealt{rinott:twostage} and the sequential KN of \citealt{kim2001fully})
and three fixed budget procedures (two-stage and sequential variations of the \OCBA of Chen and coauthors, the sequential KG variations studied by Frazier, Ryzhov, Chick, and others).

\tabref{tab:numericalprocedures} summarizes these procedures and the parameters they require in addition to the number of alternatives, $\numproj$, and the initial Stage~1 sample size for each alternative, $n_0$.

\paragraph{Parameter values of procedures.}

For \twoseosvarbound and \twoseosvarplugin, and \twoseosvarconfidence, we set the intermediate parameter $\zeta_{\numproj,n_0}$ as a function of $\numproj$ and $n_0$ based on  \citet{chick:izopcost}.
For \twoseosvarconfidence, we set $\alphavars{i}=\alphavar/\numproj$ to ensure an overall CI coverage for all variance bounds of $1-\alphavar$.

We consider two fixed-precision comparator procedures, the two-stage procedure of \cite{rinott:twostage}, and the fully sequential KN procedure of \citet{kim2001fully}.
Rinott's procedure also sets the Stage~2 sample size according to the sample variance, as we have done, but the constant of proportionality is chosen to guarantee a selection probability error rate, $\PICS \leq \alpha$, rather than to minimize expected total cost. The KN procedure is fully sequential after an initial Stage~1 set of observations. It has triangular stopping regions and also focuses on $\PICS$ guarantees.

\begin{table}[bt]
\centering
\caption{Procedures in the numerical study and their required inputs. The abbreviated procedure names used in the figures are shown in parentheses. Here $\numproj$ is the number of alternatives and $n_0$ is the initial sample size per alternative.}
\label{tab:numericalprocedures}
\footnotesize
\begin{tabular}{>{\raggedright\arraybackslash}p{0.55\linewidth}>{\raggedright\arraybackslash}p{0.08\linewidth}>{\raggedright\arraybackslash}p{0.27\linewidth}}
\hline
Procedure and implementation & Stages & Required inputs beyond $\numproj$ and $n_0$ \\
\hline
 \EOS with variance upper bounds (\twoseosvarbound), \secref{sec:freqgeneraleos}
& Two-stage
& $\boldsymbol c$, $\vecsigsqbnd$ \\
\EOS with Stage~1 CI bounds for variances (\twoseosvarconfidence), \secref{sec:freqgeneraleosunkvar}
& Two-stage
& $\boldsymbol c$, variance CI coverage $1-\alphavar$  \\
\EOS with Stage~1 sample variances (\twoseosvarplugin), \secref{sec:freqgeneraleosunkvar}
& Two-stage
& $\boldsymbol c$ \\
Sequential \EOS (\seqeos), \secref{sec:freqeosseq}
& Sequential
& $\boldsymbol c$ \\
\hline
Fixed-precision comparators \\\hline
Rinott, \citep{rinott:twostage}
& Two-stage
& $\delta$, target $\PCS$ level $1-\alpha$ \\
KN \citep{kim2001fully}
& Sequential
& $\delta$, target $\PCS$ level $1-\alpha$ \\
\hline
Fixed-budget allocation comparators \\\hline
\OCBAtwo \citep{chenhc:gradalloc96}
& Two-stage
& $\boldsymbol c$, total sampling budget $\budget$ \\
\OCBA \citep{eococba2007}
& Sequential
& $\boldsymbol c$, total sampling budget $\budget$ \\
KG \citep{frazier2008}
& Sequential & $\boldsymbol c$, total sampling budget $\budget$ \\
\hline
\end{tabular}
\end{table}

We also consider fixed-budget comparator procedures, KG and \OCBA. Each observation costs $\costs{i}$ from alternative $i$, and a total budget of $\budget$ can be exhausted, so that $\sum \costs{i} n_i \leq \budget$, with as little leftover budget as possible, on stopping. For each, a first stage of sampling is completed, and a probability distribution is computed for the unknown means and variances assuming a noninformative normal-gamma prior distribution \citep{eococba2007}.

We implement the sequential, one-step lookahead KG allocation following \citet[Section~9]{frazier2008}.
After $n_i$ observations from alternative $i$, let $\overline X_i$ and $S_i^2$ be its sample mean and sample variance, and define $\widetilde\sigma_i=\{S_i^2/[n_i(n_i+1)]\}^{1/2}$ and $z_i=-|\overline X_i-\max_{j\ne i}\overline X_j|/\widetilde\sigma_i$. Its KG factor is
\[
\mathrm{KG}_i=\widetilde\sigma_i f_{n_i-1}(z_i) / \costs{i},
\qquad
\mbox{where } f_{\nu}(z)=\frac{\nu+z^2}{\nu-1}\phi_{\nu}(z)+z\Phi_{\nu}(z),
\]
$\phi_{\nu}$ and $\Phi_{\nu}$ are the Student-$t$ density and distribution functions with $\nu$ degrees of freedom. The KG procedure sequentially samples an alternative attaining $\max_i\mathrm{KG}_i$ until the sampling budget constraint is hit; it then selects the largest sample mean. OCBA likewise receives a fixed total sample budget.  After each observation, the sequential \OCBA recomputes its plug-in target allocation and samples 1 observation at a time until the budget is exhausted.
A two-stage variation of \OCBA, called \OCBAtwo here, determines Stage~2 sample sizes using Stage~1 sample means and variances, the remaining budget after Stage~1, and standard \OCBA asymptotic allocations \citep{chenhc:tdist}.

\paragraph{Performance measures.} We will compare procedures by their expected total cost (the focus of \EOS procedures), expected sampling cost, and expected opportunity cost of selections. In addition, we explore the expected sample size and probability of correct selection.

\paragraph{Parameter values of problem configurations and implementation details.}
We tested $\numproj=2$ and $\numproj=5$ alternatives in various experiments.
Every two-stage and sequential procedure begins with the same $n_0=10$ observations from each alternative and pays for them from their budget, as do the fixed-budget \OCBA and KG procedures.
Samples were generated using independent normal observations, consistent with the assumptions of each model.

We tested several configurations of the means and variances.
Without loss of generality, the best alternative is alternative~1.
In the slippage configuration with common variances (SC-CV), $\theta_1=0$, $\theta_i=-10{,}000$ for $i=2,\ldots,5$, and $\sigma_i=100{,}000$ for every $i$. In the monotone-decreasing-means configuration with increasing variances (MDM-IV), $\theta_i=-(i-1)10{,}000$ and the standard deviations increase linearly from $\sigma_1=50{,}000$ to $\sigma_5=100{,}000$. MDM-DV uses the same means, but the standard deviations decrease linearly from $\sigma_1=100{,}000$ to $\sigma_5=50{,}000$. Thus MDM-DV places the greatest sampling uncertainty on the best alternative.

All alternatives have the same cost $\costs{i}=\cost$ per observation, with $\cost$ varied over a broad range, $\{ 2^{-4},2^{-2},\ldots,2^{6} \}$. The value of $\cost$ influences the total sample size of the \EOS and fixed-budget procedures.
The total sampling budget is $\budget=3,200$ unless otherwise specified for the fixed-budget comparator procedures.
For the Rinott and KN procedures, we used indifference-zone parameter $\delta=10{,}000$ and target $\PCS=0.90$, unless otherwise specified. These fixed-precision procedures do not consider sampling costs, and therefore their sample sizes do not depend on $\cost$ or $\budget$.

We estimate each performance measure using $1{,}000$ independent macroreplications.
For Procedure~\seqeos we cap the maximum sample size to $100{,}000$. No simulated sample paths in our experiments reached that maximum sample size.

\subsection{Penalty for Misspecifying Variance Bounds with Two-Stage \EOS Procedures}
\label{sec:numericalk2caps}

Fixed-precision indifference zone procedures are sensitive in their performance if the difference between the true means of the best two alternatives equals, is less than, or is greater than the specified mean difference $\delta > 0$ worth detecting.
In a similar spirit, we test the new two-stage \EOS procedures when the modeler specifies a standard deviation cap $\overline\sigma_{i}$ that is equal to ($\overline\sigma_{i}/\stds{i}=1$), greater than ($\overline\sigma_{i}/\stds{i}=2$), and (incorrectly) less than ($\overline\sigma_{i}/\stds{i}=1/2$) the true variance.

We study $\numproj=2$ and $\numproj=5$ alternatives. The choice $\numproj=2$ allows for a comparison with the one-stage \EOSonektwo, assuming known true variances (see Remark \ref{rem:twostochastic}).
Both \EOSonektwo and \twoseosvarbound consider unrestricted mean configurations, but differ in the variance information they use and in the objective they optimize.
We compare them for a range of costs per sample and several performance metrics on the SC-CV, MDM-IV, and MDM-DV configurations.

\begin{figure}[!tb]\centering
\includegraphics[width=\linewidth,height=0.81\textheight,keepaspectratio]{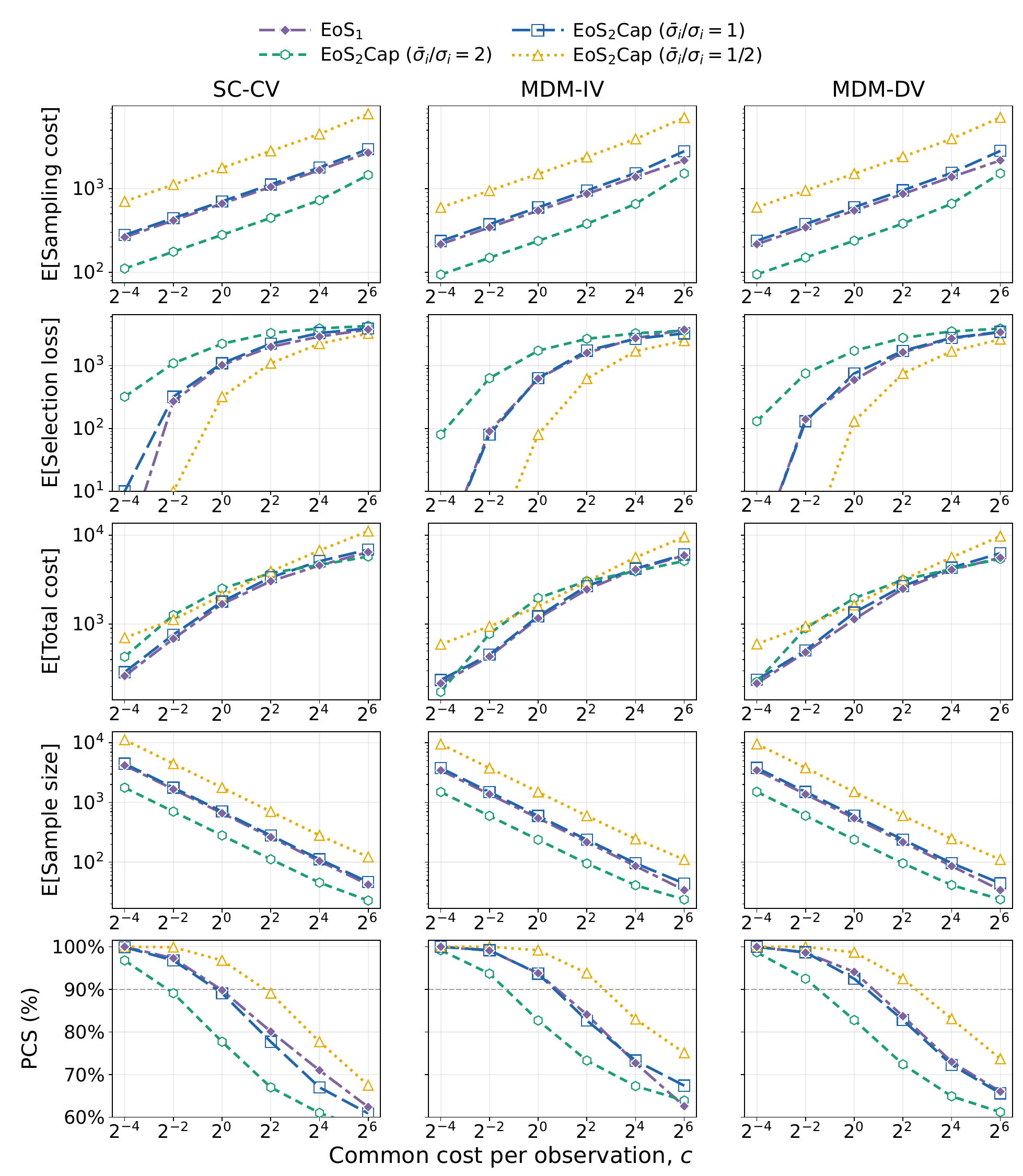}
\caption{Variance-cap sensitivity for $\numproj=2$. The \twoseosvarbound curves state the common values $\overline\sigma_i/\sigma_i\in\{1,2,1/2\}$ in that order. Rows show expected sampling cost, expected selection loss, expected total cost, expected total sample size, and PCS. Columns show SC-CV, MDM-IV, and MDM-DV configurations.}
\label{fig:k2caprobustness}
\end{figure}

\figref{fig:k2caprobustness} shows that \EOSonektwo and \twoseosvarbound with correctly specified caps, $\overline\sigma_i=\sigma_i$, perform similarly.
When the standard deviation is smaller than the cap, $\overline\sigma_i=2\sigma_i$, \twoseosvarbound selects a smaller value of $g$ because such large, conservative caps increase the worst-case expected sampling costs. It therefore samples less for each given $c$, lowering sampling cost and PCS but increasing selection loss. When the caps are invalid and underestimate the standard deviations, $\overline\sigma_i=\sigma_i/2$, \twoseosvarbound behaves in the opposite direction: it samples substantially more, pays a higher sampling cost, and achieves lower selection loss and higher PCS. Overall, the underestimated-cap version has higher total cost than \twoseosvarbound with correctly specified caps. For the conservative caps, the total-cost comparison depends on whether the sampling-cost savings offset the additional selection loss.
As an aside, sufficiently small costs $\cost$ increase sample sizes sufficiently to achieve high \PCS performance.

The results when $\numproj=5$ are largely very similar to those with $\numproj=2$ (data not shown), although \EOSonektwo can't be used when $\numproj>2$.
With five alternatives, the reduction in PCS from conservative caps is more pronounced than with two alternatives because an incorrect selection can arise from any of the four comparisons with the best alternative. In terms of total cost, the underestimated-cap version is consistently more costly than the correctly specified-cap version, whereas the effect of conservative caps depends on the configuration and the value of $\cost$.

\subsection{Does Not Knowing the Sampling Variances Hurt Two-Stage \EOS Procedures?}
\label{sec:numericalunknownvarianceq3}

The two-stage procedure \twoseosvarbound with $\overline\sigma_i=\sigma_i$ minimizes the expected total loss over all configurations of interest, but the bounds on the variances it assumes might not be known. \secref{sec:freqgeneraleosunkvar} presented two alternatives to account for unknown variances: \twoseosvarplugin uses a plugin estimator of the variances from stage~1 data, \twoseosvarconfidence uses a confidence interval for the variances.
We compare the performance of these two-stage procedures to handle uncertainty about the variances.

Specifically, we compare the perfect information setting, \twoseosvarbound with $\overline\sigma_i=\sigma_i$, with the setting \twoseosvarbound with conservatively specified bounds on variance, $\overline\sigma_i=2\sigma_i$, along with \twoseosvarplugin and with \twoseosvarconfidence using a simultaneous upper confidence of $1-\alphavar = 0.95$, so the upper confidence bound for each of the $\numproj=5$ alternatives has $\alphavars{i}=\alphavar/\numproj=0.01$.

\begin{figure}[!t]
\centering
\includegraphics[width=\linewidth,height=0.85\textheight,keepaspectratio]{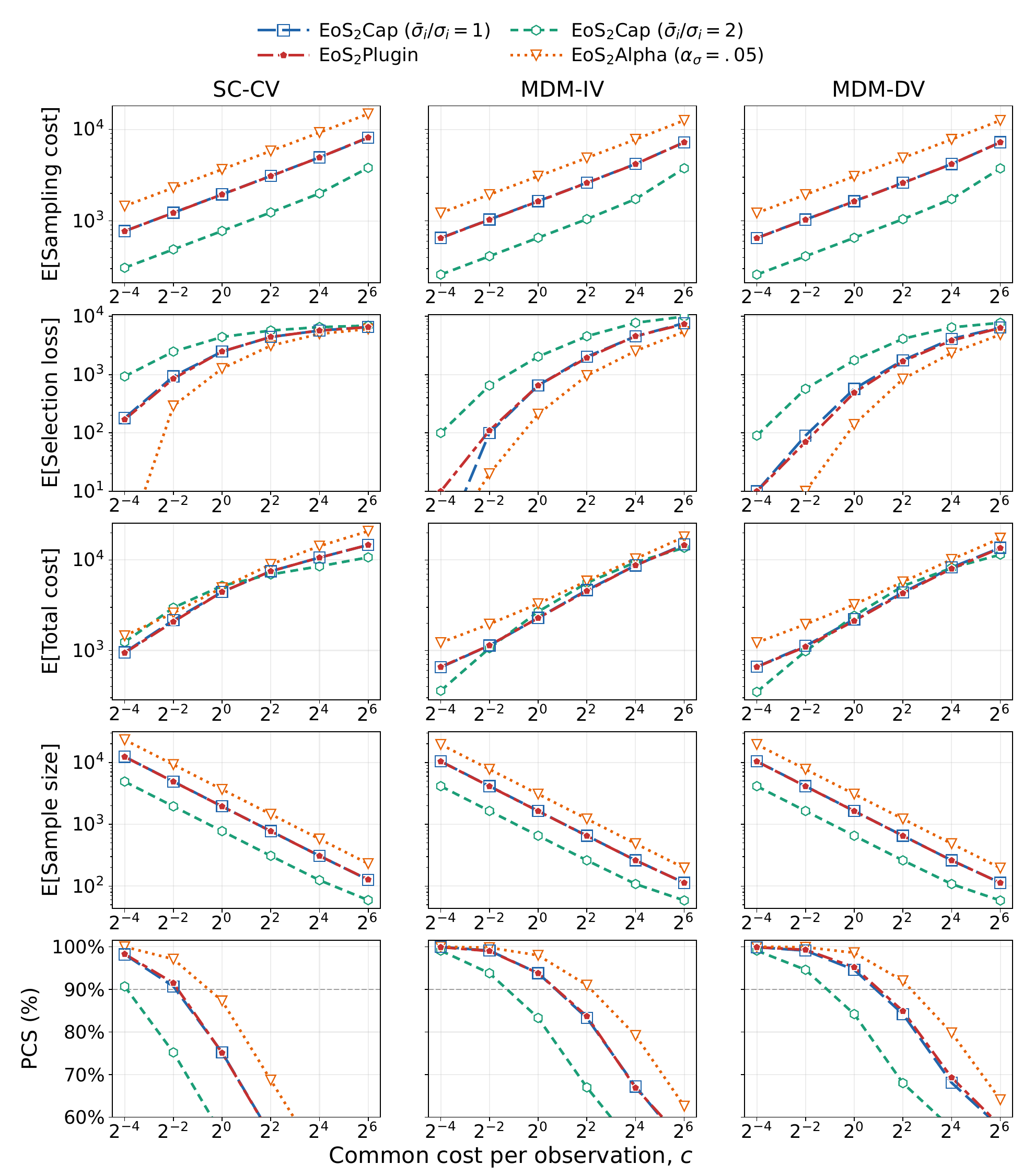}
\caption{Penalty for not knowing the sampling variances among two-stage procedures. The \twoseosvarbound references use $\overline\sigma_i/\sigma_i=1$ or $2$, and \twoseosvarconfidence uses $\alphavar=0.05$.
Rows and columns give metrics and configurations, as in \figref{fig:k2caprobustness}.}
\label{fig:twostagevarianceinfo}
\end{figure}

\figref{fig:twostagevarianceinfo} shows that \twoseosvarplugin performs very similarly to the oracle-cap benchmark, \twoseosvarbound with $\overline\sigma_i=\sigma_i$, in sample size, selection loss, total cost, and PCS. The direct substitution of the Stage~1 sample variances therefore creates little empirical penalty in the three configurations tested. \twoseosvarbound with $\overline\sigma_i=2\sigma_i$ samples less and consequently has higher selection loss and lower PCS. In contrast, \twoseosvarconfidence samples substantially more because it uses larger estimated upper confidence bounds for the variances.
This protection from the potential of higher variances lowers selection loss and raises PCS, but its higher sampling cost results in higher total cost throughout the experiments. Thus the direct plug-in performs well empirically, and did not hurt the \EOS procedure here, while the confidence-region procedure pays for its stronger protection through more sampling.

\subsection{Two-Stage EoS versus Fixed-Budget and Fixed-Precision Procedures}
\label{sec:numericaltwostagecomparison}

We now compare our proposed procedure \twoseosvarbound with loose variance bounds ($\overline\sigma_i/\sigma_i=2$) and the \twoseosvarplugin with stage~1 sample variances with two-stage comparators from the fixed precision (Rinott) and fixed-budget (\OCBAtwo) traditions.
Rinott uses $\delta=10{,}000$ and target PCS $0.90$.
\OCBAtwo uses budget $\budget=3{,}200$, including Stage~1 outcomes.All procedures use $n_0=10$, use $\numproj=5$ alternatives with the same three configurations, and the same costs.

\begin{figure}[!t]
\centering
\includegraphics[width=\linewidth,height=0.85\textheight,keepaspectratio]{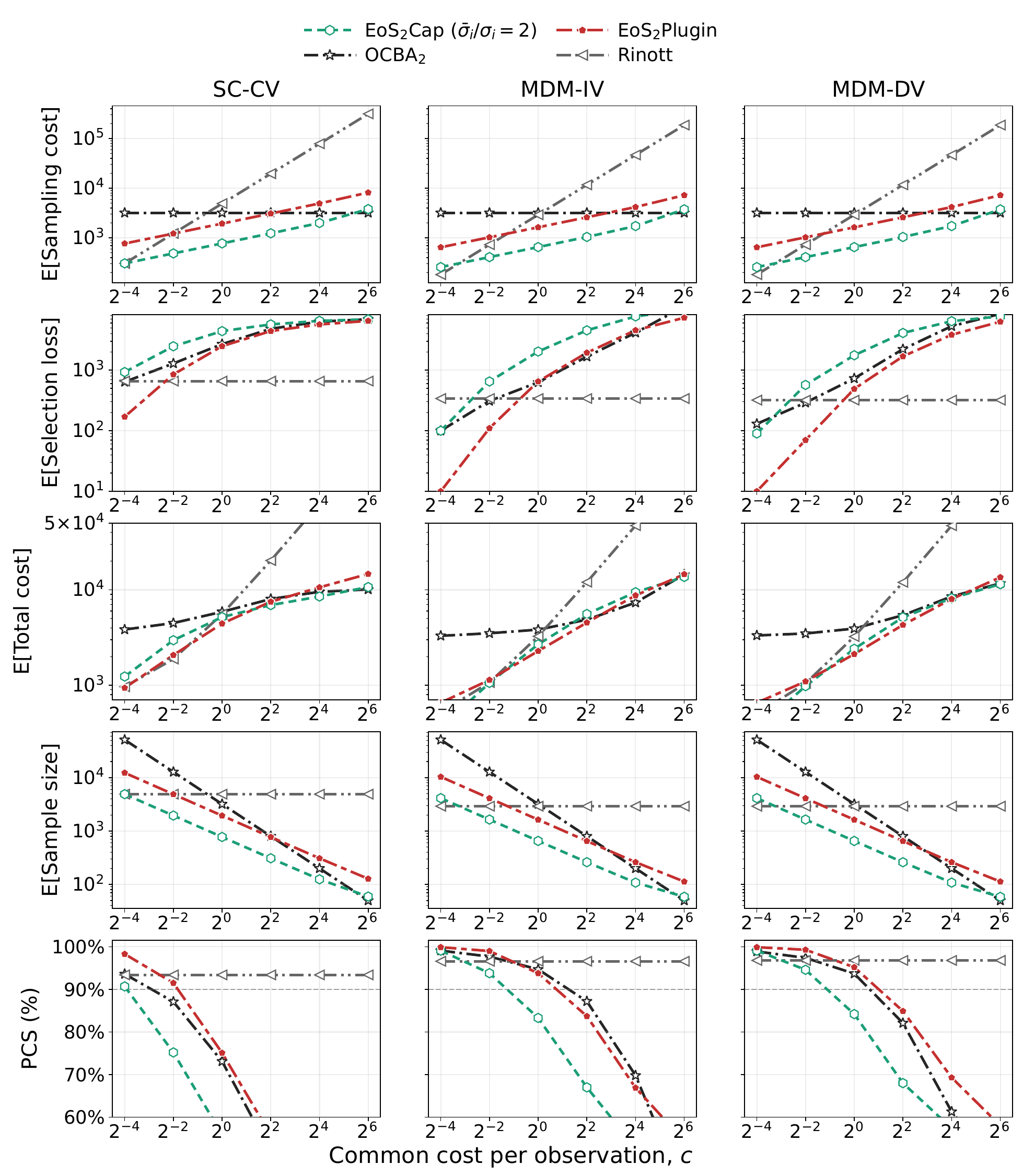}
\caption{Two-stage EoS, fixed-budget, and fixed-precision procedures for $\numproj=5$. \twoseosvarbound uses $\overline\sigma_i/\sigma_i=2$. \OCBAtwo uses sampling budget $B=3{,}200$, including Stage~1; Rinott targets $\PCS=0.90$. Rows and columns give metrics and configurations, as in \figref{fig:k2caprobustness}.}
\label{fig:twostageobjectives}
\end{figure}

\figref{fig:twostageobjectives} shows that the procedures respond differently to increases in the observation cost because they optimize different objectives.
Error bounds for total-cost curves are smaller than the size of the markers for each curve.
\twoseosvarbound and \twoseosvarplugin reduce their sample sizes as $\cost$ increases, accepting higher selection loss and lower PCS when additional observations become more expensive. Rinott maintains its fixed-precision requirement, so its sample size and PCS do not change with $\cost$, while its sampling cost increases directly with $\cost$. \OCBAtwo always exhausts the same monetary budget, so increasing $\cost$ reduces the number of observations that it can collect.

Just as different two-stage fixed-precision procedures might vary in their delivered \PICS for different configurations, no procedure has the lowest total cost in every configuration. The EoS procedures become more favorable relative to Rinott when sampling is expensive, while the comparison with \OCBAtwo depends on whether its fixed budget is appropriate for the particular configuration.

That said, \twoseosvarplugin does reasonably well across the board here, and can be recommended for two-stage procedures if the goal is minimizing total cost, rather than optimizing one of budget or selection quality subject to a constraint on the other (selection quality or budget, respectively).

\subsection{The Benefit of Sequential Sampling Relative to Two-Stages}
\label{sec:numericalsequentialbenefit}

We now ask what is gained, or lost, by shifting from a two-stage \EOS procedure, \twoseosvarplugin, to the fully sequential procedure, \seqeos, and compare these procedures with fully sequential fixed-budget (\OCBA and KG) and fixed-precision (KN) procedures.
We study $\numproj=5$ alternatives.

\begin{figure}[!t]
\centering
\includegraphics[width=\linewidth,height=0.85\textheight,keepaspectratio]{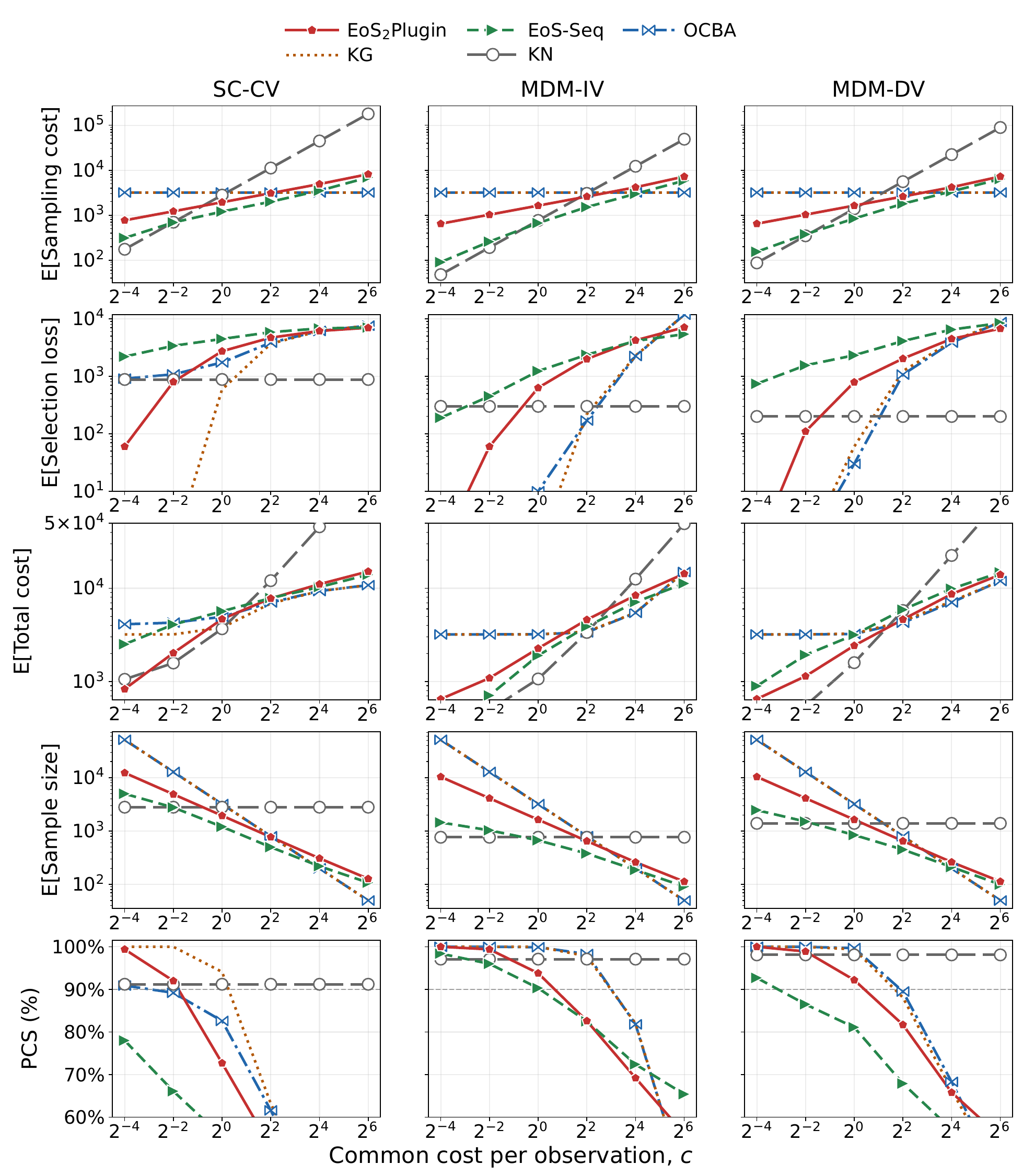}
\caption{Two-stage \twoseosvarplugin, sequential \seqeos, and fixed-budget and fixed-precision references for $\numproj=5$.
\OCBA uses monetary sampling budget $B=3{,}200$, including Stage~1, and hence uses $N(c)=\lfloor B/c\rfloor$ observations; KN targets $\PCS=0.90$.
Rows and columns give metrics and configurations, as in \figref{fig:k2caprobustness}.
}
\label{fig:sequentialbenefit}
\end{figure}

\figref{fig:sequentialbenefit} shows that the benefit of \seqeos depends strongly on the configuration. Compared with \twoseosvarplugin, \seqeos performs well in MDM-IV, where it uses the estimated mean gaps to avoid continued sampling from alternatives that appear clearly inferior. In SC-CV, it is more costly when sampling is inexpensive but becomes favorable at the largest values of $\cost$. In MDM-DV, where the best alternative has the largest variance, the sampling savings from \seqeos do not compensate for its higher selection loss, and its total cost is higher throughout the experiment. These results show that this specific choice of sequential updating can reduce total cost when the estimated gaps reliably identify inferior alternatives, but it can also stop too early and result in high selection costs.

We explore this issue further. When the cost per observation is high, the total cost of KN is worse than that of \seqeos and \twoseosvarplugin.
When the cost per observation is very low, the parameters chosen for KN can lead to better total costs than those of \seqeos, even though the total sample size of KN does not depend on the cost per observation. In that setting, \seqeos stops too early, resulting in a high selection losses. This is not surprising: the greedy, gradient-based nature of the stopping rule of \seqeos makes it similar to the one-step lookahead $KG_{1}$ index, which is useful for allocation rules but is known to be poor for stopping rules \citep{chick:12a}. A multi-step (and potentially multi-arm) lookahead stopping rule to extend \seqeos would be expected to give rise to superior \EOS performance for total cost, and we leave this for future research.

\section{Discussion}
\label{sec:discuss}

Taxonomies can help to structure ideas, cross-fertilize research, and identify strengths and weaknesses of different methods for different applications. \RnS, BAI, and related discrete stochastic optimization taxonomies that classify procedures exclusively as fixed-budget or fixed-precision are incomplete. They do not account for a broad class of procedures developed decades ago and initially called Bayesian decision theoretic designs \citep{chernoff:sequential61,degroot:decisions}, that are now called economics of selection (\EOS) procedures in the WSC community \citep{chick:09a,chick:12a} and value-based designs in the clinical trial and health economics community \citep{chick2017bayesian,ENACTcactus,ENACTposition}.

We used such cross-fertilization to propose a new frequentist IZ \EOS procedure in \secref{sec:freqeos}. We demonstrated that \EOS is not a Bayesian versus frequentist concept, nor is it a contrast between fixed-sample size or response adaptive, or an asymptotic versus finite sample validity (e.g., \citealp{degroot:decisions} studies $\cost \rightarrow 0$ for a large $\stoptimeopt$ analysis, \citealp{lai:review2001} notes many related asymptotic results).
\EOS jointly considers sampling costs and the penalty from incorrect selections together, rather than optimizing just one or the other.
Other taxonomies are useful too \citep{raghuetal15constraint,hongfanluo}, but have not been covered here, e.g., distribution of outcomes, side constraints or not.

More areas of cross-fertilization are evident. For example, there are Bayesian fixed-budget allocation rules to optimize \EOC for multiple arms \citep{CGY2021Multiarm} or for contextual \RnS \citep{AlbChiZoum:25predprogMS}, that were not presented with \EOS stopping rules. The frequentist approach to fixed-step lookaheads for multiple arms in \secref{sec:freqeoskbig} combined with frequentist stopping criteria in \secref{sec:freqeosseq} motivates new Bayesian lookahead stopping rules that generalize the KG$^\ast$ optimal lookahead for one arm, and may prove useful in multi-arm and contextual \RnS and clinical trial design.

According to \citet{bather:chernoffconversation}, Chernoff found initial sequential sampling approaches to hypothesis testing based on the IZ framework to be ``rather primitive'', that ``the proper decision theoretic formulation was to take the cost of making the wrong decision into account'', and that it is ``necessary to introduce the cost of sampling and then optimize with respect to the Bayesian criterion.'' Whether one agrees with all of Chernoff's assertions or not, it seems hard to argue that the choice of \PICS threshold $\alpha$ should be independent of the cost of sampling or the financial impact of a selection decision.
If fixed-precision is about hypothesis testing, and fixed-budget is about resource consumption, then \EOS is about expected profit. In this paper, we propose and analyze the first two frequentist \EOS procedures that account for $\numproj\geq 2$ alternatives and potentially unknown variances, and propose the first such frequentist sequential heuristic. Thus, \EOS is not exclusively Bayesian.

Financial value is central to \RnS applications in business and health.
If operations focuses on cost reduction (respecting a budget) and marketing focuses on increasing revenue generation, then perhaps fixed-budget is like operations and marketing is like fixed-precision. Economics of selection procedures maximize profit from an \RnS project, but optimize neither revenue nor cost alone.
The \EOS approach has been proposed for use in so-called value-adaptive clinical trials: adaptive because of response adaptivity, and value-based in that the health-economic cost-effectiveness of health technology adoption decisions is balanced with the cost of the clinical trial. \citet{ENACTcactus} and \citet{ENACTposition} assess what would have happened had three UK-funded clinical trials been run with the approach in \secref{sub:Bayesseqexample}.
Their response gives one practical way of ``baking in'' the cost of clinical research into the assessment of the real-world value of clinical research projects, as called for by the \citet{NHSEnglandbakein}.

We encourage the simulation community to consider the full palette of approaches to \RnS, including \EOS. There are exciting opportunities for new work building on \secrefb{sec:freqeos}{sec:freqeosseq}, including:
new frequentist \EOS procedures; better or new stopping rules to better balance sampling costs and expected selection losses for more output distributions or for multi-arm trials; derivations of frequentist guarantees for Bayesian sequential procedures, both fixed-budget and \EOS.
One might explore non-additive utility as a function of sampling cost and selection quality for \EOS in \eqref{def:econselect}.

\ACKNOWLEDGMENT{Chick and Li thank Dr.~Simba Gill and Sabi Dau for their generous support of INSEAD's Healthcare Management Initiative. Hong thanks INSEAD for hosting his academic visit in March 2026, which contributed to the development of this article. We thank Barry Nelson for feedback on an earlier draft.}

\bibliographystyle{informs2014}
\bibliography{ref}

\begin{APPENDICES}
\section{Mathematical Proofs and Auxiliary Results for the Main Paper}
\label{app:section4aux}
\label{app:section5proofs}
We prove the mathematical claims in the main paper, subsection by subsection.

\subsection{\secref{sec:freqknownstandard}: One alternative vs a known standard, known variance, one stage}

We first prove the two informally stated claims in \secref{sec:freqknownstandard}, then prove the formally stated claims there. For $\theta>0$, the reduction in expected selection loss obtained by increasing the sample size from one to two is
\[
\{R(1,\theta)-\cost\}-\{R(2,\theta)-2\cost\}
=\theta\left[
\Phi\left(-\frac{\theta}{\sigma}\right)
-\Phi\left(-\frac{\theta\sqrt{2}}{\sigma}\right)
\right].
\]
The second observation costs an additional $\cost$, so the associated reduction in expected total cost is
\[
R(1,\theta)-R(2,\theta)
=\theta\left[
\Phi\left(-\frac{\theta}{\sigma}\right)
-\Phi\left(-\frac{\theta\sqrt{2}}{\sigma}\right)
\right]-\cost.
\]
Therefore $R(2,\theta)<R(1,\theta)$ if and only if the reduction in expected selection loss exceeds the additional sampling cost $\cost$. Define
\begin{equation}
\label{eq:onestagecriticalcost}
\cost_{\rm crit}^{(1)}(\sigma)
=\sup_{\theta>0}\theta\left[
\Phi\left(-\frac{\theta}{\sigma}\right)
-\Phi\left(-\frac{\theta\sqrt{2}}{\sigma}\right)
\right].
\end{equation}
Thus $\cost_{\rm crit}^{(1)}(\sigma)$ is the largest reduction in expected selection loss obtained by increasing the sample size from one to two over all configurations of the mean.

\begin{proposition}[No uniformly pointwise-optimal one-stage sample size]
\label{prop:nooneStagepointwise}
The constant in \eqref{eq:onestagecriticalcost} is strictly positive and finite. If $0<\cost<\cost_{\rm crit}^{(1)}(\sigma)$, no positive integer sample size minimizes \eqref{eq:totalIZcost} at every $\theta\in\mathbb R$. If sampling zero observations is allowed and the procedure selects each alternative with probability $1/2$ when no data are available, no uniformly pointwise-optimal sample size exists for any finite $\cost>0$.
\end{proposition}

\proof{\textbf{Proof of Proposition~\ref{prop:nooneStagepointwise}.}}
At $\theta=0$, the selection loss is zero and $R(n,0)=\cost n$, so any uniformly pointwise-optimal positive integer sample size would have to equal $1$. The function inside the supremum in \eqref{eq:onestagecriticalcost} is continuous and strictly positive for $\theta>0$. It tends to zero as $\theta\downarrow0$ because it is bounded above by $\theta/2$. As $\theta\to\infty$, Mills' inequality gives
\[
0\leq \theta\left[
\Phi\left(-\frac{\theta}{\sigma}\right)
-\Phi\left(-\frac{\theta\sqrt{2}}{\sigma}\right)
\right]
\leq \sigma\phi\left(\frac{\theta}{\sigma}\right)
\longrightarrow0.
\]
Consequently, the supremum is positive, finite, and attained at an interior point. If $\cost$ is below that supremum, then for some $\theta>0$,
\[
R(2,\theta)-R(1,\theta)
=\cost-\theta\left[
\Phi\left(-\frac{\theta}{\sigma}\right)
-\Phi\left(-\frac{\theta\sqrt{2}}{\sigma}\right)
\right]<0,
\]
contradicting the required optimality of $n=1$. If sampling zero observations is allowed, selecting each alternative with probability $1/2$ without data gives $R(0,\theta)=\theta/2$ for $\theta>0$. Taking one observation and selecting according to its sign gives $R(1,\theta)=\cost+\theta\Phi(-\theta/\sigma)$. Hence
\[
R(1,\theta)-R(0,\theta)
=\cost-\theta\left\{\frac12-\Phi(-\theta/\sigma)\right\}.
\]
The quantity $\theta\{1/2-\Phi(-\theta/\sigma)\}$ diverges to infinity as $\theta\to\infty$, so this difference is negative for some $\theta$ for every finite sampling cost.\halmos
\endproof

The following elementary result supplies the inner-maximization step used in \lemref{lem:IZk2}.

\begin{lemma}[Shape of the normal-tail loss]
\label{lem:normaltailshape}
Let $f(z)=z\Phi(-z)$ for $z\geq0$. There is a unique $z^*\in(0,1)$ satisfying
\begin{equation}
\label{eq:appendixzstarroot}
\Phi(-z^*)=z^*\phi(z^*).
\end{equation}
The function $f$ is strictly increasing on $[0,z^*]$ and strictly decreasing on $[z^*,\infty)$. Consequently, $z^*$ is the unique maximizer of $f$ on $[0,\infty)$.
\end{lemma}

\proof{Proof.}
Differentiate $f$ and write
\[
h(z)=f'(z)=\Phi(-z)-z\phi(z).
\]
Then $h(0)=1/2$ and $h'(z)=(z^2-2)\phi(z)$. Thus $h$ is strictly decreasing on $(0,\sqrt2)$. Moreover, Mills' inequality gives $\Phi(-z)<\phi(z)/z$ for every $z>0$, so for $z\geq1$,
\[
h(z)<\phi(z)\left(\frac1z-z\right)\leq0.
\]
In particular, $h(0)>0$ and $h(1)<0$. Continuity and strict decrease on $(0,1)$ therefore give exactly one zero $z^*\in(0,1)$, and this zero satisfies \eqref{eq:appendixzstarroot}. The same Mills bound gives $h(z)<0$ for all $z\geq1$, ruling out any further zero. Hence $h(z)>0$ for $0\leq z<z^*$ and $h(z)<0$ for $z>z^*$, which proves the claimed monotonicity and uniqueness of the maximizer.\halmos
\endproof

\proof{\textbf{Proof of \lemref{lem:IZk2}.}}
For fixed $n$, the sampling-cost term $\cost n$ does not depend on $\theta$. Moreover, the selection-loss term $|\theta|\Phi(-|\theta|\sqrt n/\sigma)$ is symmetric in $\theta$. Therefore, the inner supremum in \eqref{def:econselectIZ} can be computed by maximizing the selection loss over $\theta\in[\delta,\infty)$. Under the one-to-one change of variables $z=\theta\sqrt n/\sigma$, this problem is
\[
\sup_{\theta\geq\delta}
\theta\Phi(-\theta\sqrt n/\sigma)
=\frac{\sigma}{\sqrt n}
\sup_{z\geq \delta\sqrt n/\sigma} f(z).
\]
Let $a_n=\delta\sqrt n/\sigma$ denote the lower endpoint of the feasible set for $z$. By \lemref{lem:normaltailshape}, if $a_n\leq z^*$, then the feasible set contains the unique global maximizer $z^*$, so the maximizing value is $z^*$. If $a_n>z^*$, then every feasible $z$ lies in the region where $f$ is strictly decreasing, so the maximum is attained at the left endpoint $a_n$. Hence the value of $z$ that maximizes the selection loss is
\[
z_{\rm LFC}(n)
 =\max\left\{\frac{\delta\sqrt n}{\sigma},z^*\right\}.
\]
Mapping $z_{\rm LFC}(n)$ back to the mean scale and restoring the negative, symmetric configuration gives the two LFCs in \eqref{eq:thetaLFC}.

For $\delta>0$, the condition $a_n\leq z^*$ is equivalent to $n\leq n^+_\delta$. In that case, the supremum of the selection loss is $\sigma f^*/\sqrt n$. When $n\geq n^+_\delta$, the maximizing mean gap is $|\theta|=\delta$, and the supremum is $\delta\Phi(-\delta\sqrt n/\sigma)$. Adding the common sampling cost $\cost n$ gives the two conditions in \eqref{eq:freqriskpiecewise}. They agree at $n^+_\delta$ because $f^*/z^*=\Phi(-z^*)$. When $\delta=0$, $a_n=0<z^*$ for every $n>0$, so the first condition applies throughout.\halmos
\endproof

\proof{\textbf{Proof of \thmref{thm:freqoptimaln}.}}
Let $J_{\rm low}(n)=\cost n+\sigma f^*n^{-1/2}$ denote the branch of $J_\delta$ for $0<n\leq n^+_\delta$ and, for $\delta>0$, let $J_{\rm high}(n)=\cost n+\delta\Phi(-d)$ denote the branch for $n\geq n^+_\delta$, where $d=\delta\sqrt n/\sigma$. Their derivatives are
\[
\begin{aligned}
J_{\rm low}'(n)&=\cost-\frac{\sigma f^*}{2n^{3/2}},
&J_{\rm low}''(n)&=\frac{3\sigma f^*}{4n^{5/2}}>0,\\
J_{\rm high}'(n)&=\cost-\frac{\delta^3\phi(d)}{2\sigma^2d},
&J_{\rm high}''(n)&=\frac{\delta^5\phi(d)(1+d^2)}{4\sigma^4d^3}>0.
\end{aligned}
\]
Thus each applicable branch is strictly convex. When $\delta=0$, $n^+_\delta=\infty$, $J_\delta=J_{\rm low}$ for every $n>0$ and $\cost_\delta=0$; because $J_{\rm low}$ diverges as $n\downarrow0$ and as $n\to\infty$, the zero of $J_{\rm low}'$ is its unique minimizer, $n_{\rm low}=(\sigma f^*/(2\cost))^{2/3}$, giving the first expression in \eqref{eq:freqoptimaln}.

Now suppose $\delta>0$. At $n=n^+_\delta$, we have $d=z^*$ and $f^*=z^*\Phi(-z^*)=(z^*)^2\phi(z^*)$, so the two branches have the same value and the same derivative, $\cost-\cost_\delta$, at their junction. The positive second derivatives above then imply that $J_\delta$ is continuously differentiable and strictly convex, and its divergence as $n\downarrow0$ and as $n\to\infty$ gives a unique minimizer. The stationary point $n_{\rm low}$ lies in the low-sample-size branch exactly when $n_{\rm low}\leq n^+_\delta$, equivalently $\cost\geq\cost_\delta$, which gives the first expression in \eqref{eq:freqoptimaln}. If $0<\cost<\cost_\delta$, the derivative at $n^+_\delta$ is negative, so the unique minimizer lies in the high-sample-size branch and satisfies
\begin{equation}
\label{eq:largefreqfoc}
\cost=\frac{\delta^3\phi(d)}{2\sigma^2d}.
\end{equation}

The function $\phi(d)/d$ is strictly decreasing for $d>0$ because its derivative is $-\phi(d)(1+d^{-2})<0$, so the positive root is unique. Substituting $\phi(d)=(2\pi)^{-1/2}e^{-d^2/2}$ into \eqref{eq:largefreqfoc} and squaring two positive quantities gives
\[
d^2e^{d^2}=\frac{\delta^6}{8\pi\cost^2\sigma^4}.
\]
Because the right-hand side is positive and $d^2\geq0$, inversion of $w\mapsto we^w$ on $[0,\infty)$ gives $d^2=W_0(\delta^6/(8\pi\cost^2\sigma^4))$; no other real branch yields a nonnegative solution at this positive argument. Substitution into $n=\sigma^2d^2/\delta^2$ gives the second line of \eqref{eq:freqoptimaln}, while at $\cost=\cost_\delta$ both formulas give $n^*=n^+_\delta$. Finally, imposing $n\geq1$ changes the continuous minimizer to $\bar n^*$, and strict convexity implies that the integer optimum is obtained by comparing $J_\delta(\lfloor\bar n^*\rfloor)$ and $J_\delta(\lceil\bar n^*\rceil)$.\halmos
\endproof

\subsection{\secref{sec:freqeoskbig}: Two stages, multiple alternatives, potentially unknown variances}

\proof{\textbf{Proof of Proposition~\ref{prop:expectedtwostagesamples}.}}
Write $Y_i=\nu S_i^2/\sigma_i^2\sim\chi_\nu^2$. The event $g^2S_i^2\leq n_0$ is $Y_i\leq a_i(g)$, so
\[
\E[\widetilde N_i(g)]
=n_0F_\nu(a_i(g))
+\frac{g^2\sigma_i^2}{\nu}
\E\!\left[Y_i\indic\{Y_i>a_i(g)\}\right].
\]
The chi-squared densities satisfy $y f_\nu(y)=\nu f_{\nu+2}(y)$; hence the truncated expectation is $\nu\overline F_{\nu+2}(a_i(g))$, proving \eqref{eq:exactexpectedN}. Because $S_i^2$ has a continuous distribution, the breakpoint event has probability zero. On every compact interval of positive $g$, the pointwise derivative is dominated by a constant multiple of the integrable variable $S_i^2$. Dominated differentiation therefore gives
\[
\frac{d}{dg}\E[\max\{n_0,g^2S_i^2\}]
=\E[2gS_i^2\indic\{g^2S_i^2>n_0\}],
\]
which is \eqref{eq:exactexpectedNderivative}. Similarly, using $S_i^2=\sigma_i^2Y_i/\nu$ and differentiating with respect to $\sigma_i^2$ gives
\[
\frac{\partial}{\partial\sigma_i^2}\E[\widetilde N_i(g)]
=\frac{g^2}{\nu}\E[Y_i\indic\{Y_i>a_i(g)\}]
=g^2\overline F_{\nu+2}(a_i(g)),
\]
which is strictly positive and proves \eqref{eq:exactexpectedNvariancederivative}. Finally, the ceiling in \eqref{eq:twostagesamples} gives a pointwise bound $0\leq N_i(g)-\widetilde N_i(g)<1$, a result stronger than the claimed bound on expected sample sizes.\halmos
\endproof

\proof{\textbf{Proof of Proposition~\ref{prop:frequnbounded}.}}
Because $\widetilde N_i(g)\geq g^2S_i^2$ and $\E[S_i^2]=\sigma_i^2$, $\E[\widetilde N_i(g)]\geq g^2\sigma_i^2$, which diverges as $\sigma_i^2\to\infty$. Moreover, $N_i(g)\geq\widetilde N_i(g)$ and the selection loss is nonnegative, so the actual worst-case expected total cost diverges as well.\halmos
\endproof

\proof{\textbf{Proof of \propref{prop:unknownvarchirefinement}.}}
Differentiating \eqref{eq:exactexpectedNderivative} once more gives
\[
\frac{d^2}{dg^2}\E[\widetilde N_i(g)]
=2\sigma_i^2\left\{
\overline F_{\nu+2}(a_i(g))+2a_i(g)f_{\nu+2}(a_i(g))
\right\}>0.
\]
Consequently, $\widetilde H$ is strictly convex. Its derivative tends to $-\infty$ as $g\downarrow0$ and is positive for all sufficiently large $g$, so it has a unique zero. Setting that derivative to zero yields \eqref{eq:tightergFOC}.\halmos
\endproof

\textit{\corref{cor:boundedvar} and \corref{cor:knownvar}} follow directly from \propref{prop:unknownvarchirefinement}, the proof by \citet{chick:izopcost} that Procedure $OC_f(g,n_0)$ delivers an $\EOC \leq \EOCmax$ guarantee, and the framing of frequentist \EOS procedures as minimizing, over a set of problem configurations of interest, a balance between sampling costs and selection losses.

\end{APPENDICES}

\end{document}